\documentclass[usenatbib]{mn2e}
\usepackage{graphicx}
\usepackage{dcolumn}
\usepackage{subfigure}

   \author[R. Oechslin, H.-T. Janka]{R. Oechslin, H.-T. Janka\\
   Max-Planck-Institut f\"ur Astrophysik, Karl-Schwarzschild-Str. 1, D-85741 Garching, Germany}

   \title[Torus Formation in Neutron Star Mergers]{Torus Formation in
   Neutron Star Mergers and Well-Localized Short Gamma-Ray Bursts}
   \date{}

\begin{document}
   \maketitle

   \begin{abstract} 

   Merging neutron stars (NSs) are hot candidates for the still enigmatic
sources of short gamma-ray bursts (GRBs). If the central engines of the
huge energy release are accreting relic black holes (BHs) of such mergers,
it is important to understand how the properties of the 
BH-torus systems, in particular disc masses and mass and
rotation rate of the compact remnant, are linked to the
characterizing parameters of the NS binaries. For this purpose we
present relativistic smoothed particle hydrodynamics simulations
with conformally flat approximation of the Einstein field equations
and a physical, non-zero temperature equation of state. Thick
disc formation is highlighted as a dynamical process caused by
angular momentum transfer through tidal torques during the merging
process of asymmetric systems or in the rapidly spinning triaxial 
post-merger object. Our simulations support the possibility that the 
first well-localized short and hard GRBs~050509b, 050709,
050724, 050813 have originated from NS merger events and are powered
by neutrino-antineutrino annihilation around a relic BH-torus system. 
Using model parameters based on this assumption,
we show that the measured GRB energies and durations lead to estimates
for the accreted masses and BH mass accretion rates which are 
compatible with theoretical expectations.
In particular, the low energy output and short duration of GRB~050509b
set a very strict upper limit of less than 100$\,$ms for the time 
interval after the merging until the merger remnant has collapsed 
to a BH, leaving an accretion torus with a small mass of only
$\sim 0.01\,M_\odot$. This favors a (nearly) symmetric NS+NS binary 
with a typical mass as progenitor system.

   \end{abstract}

\section{Introduction}

Merger events of NS+NS or NS+BH binaries do not only belong to the
strongest known sources of gravitational wave (GW) radiation, they are
also widely favored as origin of the class of short, hard 
GRBs (\citealt{blinnikov1984, paczynski1986, eichler1989,
paczynski1991, narayan1992}).
The recent first good localizations of short bursts, 
GRB~050509b, 050709, 050724, 050813, by the {\em Swift} and {\em Hete} 
satellites at redshifts between 0.160 and 0.722 (see \citealt{fox2005},
\citealt{gehrels2005}, \citealt{villasenor2005}, and references therein)
were interpreted as a possible confirmation of this hypothesis
(\citealt{fox2005, bloom2005, hjorth2005a,hjorth2005b, lee2005}), because
the bursts have observational characteristics which are different from
those of long GRBs, but which are in agreement with expectations for compact object mergers.

The central engines of such bursts are still poorly understood
and observationally undetermined. But it seems unlikely that
the energies required for typical short GRBs are set free during
the dynamical phase of the merging of two NSs (
\citealt{ruffert1996}). The production of GRBs by the neutrino
emitting, hot post-merger NS is also disfavored, because the
high mass loss rates in a neutrino-driven wind, which is caused 
by neutrino energy deposition near 
the NS surface, rules out the production of 
high Lorentz factor outflow (\citealt{woosley1992}, 
\citealt{woosley1993}). Instead, the long-time accretion of a
BH formed from a transiently stable, supramassive or 
hypermassive merger remnant (for definitions of these terms, see \citealt{morrison2004}) is a much more promising
source (e.g., \citealt{woosley1993b}, \citealt{ruffert1999}, \citealt{popham1999}, \citealt{rosswog2003b}, \citealt{lee2005b}), provided the
BH is surrounded by a sufficiently massive accretion torus. 
Due to the geometry of the BH-torus system with relatively
baryon-poor regions along the rotation axis, thermal energy 
release preferentially above the poles of the BH by the 
annihilation of neutrino-antineutrino ($\nu\bar\nu$) pairs 
(\citealt{jaroszynski1993}, \citealt{mochkovitch1993}) can lead to collimated, 
highly relativistic jets of baryonic matter with properties in 
agreement with those needed to explain short GRBs
\citep{aloy2005}. Ultrarelativistic jets were found to develop 
if the rate of thermal energy
deposition per unit solid angle is sufficiently large.
Alternatively or in addition, magneto-hydrodynamic (MHD) 
field amplification in the rapidly spinning disc could drive polar
jets, or magnetic coupling between the rotating BH and the
girding accretion torus could tap the rotational energy of 
the BH \citep{blandfordznajek} and could help powering MHD-driven
outflow (e.g., \citealt{brown2000}, \citealt{drenkhahn2002}).

Torus mass as well as BH mass and rotation are thus crucial 
parameters that determine the energy release from the merger 
remnant on the secular timescale of the viscous evolution 
of the BH-torus system. In this letter we present a first
set of results from new three-dimensional (3D), relativistic
smoothed particle hydrodynamics simulations of NS
mergers with a physical equation of state (EoS), which aim at
establishing the link between the remnant properties and those
of the binary systems. Previous such attempts (see 
\citealt{baumgarte2003} for a review) were either performed in the Newtonian limit (e.g., 
\citealt{ruffert2001} and references therein;  \citealt{rosswog2003b}), or with polytropic or otherwise
radically simplified treatments of the high- and low-density
EoSs and their temperature-dependent behaviour 
(e.g., \citealt{shibata2003}, \citealt{shibata2005}), or 
they considered disc formation by
viscous or magnetic effects during the secular evolution 
of hypermassive NSs (\citealt{duez2004}, \citealt{duez2005}). 
Here we instead highlight torus formation 
in NS mergers as a dynamical process and argue that this is consistent
with the recent observations of short GRBs.

\section{Numerical Methods and Models}

We employ an improved version of our relativistic smoothed particle
hydrodynamics (SPH) code (\citealt{oechslin2002}, \citealt{oechslin2004}), which solves the
relativistic hydrodynamics equations together with the Einstein field
equation in the conformally flat approximation (\citealt{isenberg1980},
\citealt{wilson1996}). The code now allows for the use of a tabulated,
non-zero temperature EoS and solves the energy equation in a form
without explicit time derivatives of the metric elements on the RHS
(cf. \citealt{oechslin2002}, eqn. A8). The properties of the stellar
plasma are described by the non-zero temperature EoS of
\citealt{shen1998, shen1998b}, which is used with baryonic density, internal energy and electron fraction (proton-to-baryon ratio) $Y_e$ as input,
and pressure and temperature as output. Since the backreaction of the neutrino emission on the stellar fluid is small on the timescales considered here, we ignore neutrinos in our models. 
As a consequence, the electron fraction $Y_e$ needed by the EoS
as an input remains constant on the Lagrangian SPH
particles. Details on the numerical scheme will be presented in \citet{oechslin2006b}

We start our simulations shortly before the tidal instability
sets in and follow the evolution with typically 400'000 SPH particles through merging and torus
formation until either the collapse of the merger remnant sets in, or a quasi-stationary
state has formed.

Initial conditions for our simulations are generated by placing two NSs in hydrostatic equilibrium on a
circular orbit of a given orbital distance and by relaxing the
configuration with the use of a small damping force into a circular orbital motion with an irrotational
spin state. The orbital velocity is adjusted during this process in
order to obtain the binary in orbital equilibrium. The initial electron fraction
$Y_e$ is obtained by requiring neutrinoless $\beta$-equilibrium
for cold neutron star matter, i.e., we determine $Y_e$ from the
condition that the (electron) neutrino chemical potential $\mu_{\nu_e}$
is equal to zero for a given density (and initial temperature $T\sim 0$).

We consider a variety of models with different NS masses and mass
ratios. Other unknown parameters like the NS spins and the EoS are kept
fixed except for two models where thermal effects in the EoS are neglected.
This is realized by reducing the EoS to the twodimensional slice at
$T=0$ with density and $Y_e$ as input and pressure and
internal energy as output. This reduction has no influence as long as
shocks are absent, i.e. as long as the fluid evolves
adiabatically. In the presence of shocks, however, the $T=0$ case 
corresponds to the extreme situation that a very efficient cooling 
mechanism extracts immediately the entropy and internal energy 
generated in shocks.

In Table \ref{tab:inittable}, the key parameters
characterizing our different models are summarized. We have investigated
NS+NS binaries with mass ratios $q$ between 0.55 and unity, roughly
covering the theoretical range obtained by stellar population synthesis
calculations \citep{bulik2003}. Predictions of the mass distribution,
however, are sensitive to parameters that govern single and binary star
evolution and still contain significant uncertainties. A comparision
with observations of galactic double NS binaries, which currently
span a range of $q$-values between 0.67 and nearly unity (see the
recent reviews by \citealt{lattimer2004} and \citealt{stairs2004}), is therefore difficult and is also hampered by the still small number of
detected NS+NS systems.

In the following analysis and discussion of our models, we have 
synchronized the time axis to the time of maximal gravitational 
wave luminosity. This allows for a better comparison of the temporal 
evolution.

\begin{table*}
\begin{center}
\caption{Parameters of the considered NS+NS models. $M_1$ and $M_2$
denote the individual gravitational masses of the NSs in isolation, while $M_{\mathrm{sum}}=M_1+M_2$
stands for the sum of the two. Note that the total
gravitational mass $M$ is slightly smaller than $M_{\mathrm{sum}}$ because $M$
also involves the negative gravitational binding energy between the two stars.
$M_0$ is the total baryonic mass, $q=M_1/M_2$ and
$q_{\mathrm{M}}=M_{0,1}/M_{0,2}$ are the gravitational and baryonic
mass ratios, respectively. `Shen' stands for the full, non-zero
temperature EoS of \citealt{shen1998,shen1998b}, while `Shen\_c' denotes the $T=0$ slice of this EoS table. Characteristic data of the post-merger system are read off at $t\sim 6$ms. $M_{\mathrm{0,disc}}$ and $J_{\mathrm{disc}}$ are the
baryonic mass and the angular momentum of the disc,
respectively. $M_{\mathrm{rem}}$ is the remnant gravitational mass,
and $a_{\mathrm{rem}}=(J_{\mathrm{total}}-J_{\mathrm{disc}})/M_{\mathrm{rem}}^2$ is the corresponding spin
parameter. $j_{\mathrm{ISCO}}$ and $r_{\mathrm{ISCO}}$ are the specific angular momentum and the radius of the ISCO of a Kerr-BH with the gravitational mass and spin parameter of
the remnant.}
\label{tab:inittable}
\begin{tabular}{c|c|c|c|c|c|c|c||c|c|c|c|c|c}
Model & $M_1$ & $M_2$ & $M_{\mathrm{sum}}$ & $M_0$ & $q$  & $q_{\mathrm{M}}$ &EoS& $M_{\mathrm{0,disc}} $& $J_{\mathrm{disc}}$&$M_{\mathrm{rem}}$ & $a_{\mathrm{rem}}$&$j_{\mathrm{ISCO}}$&$r_{\mathrm{ISCO}}$\\
\hline
Unit & $M_\odot$&$M_\odot$&$M_\odot$&$M_\odot$& & & &$M_\odot$&$\times 10^{48}\mathrm{g cm}^2$/s&$M_\odot$& &$\times 10^{16} \mathrm{cm}^2/$s&km\\
\hline
\hline

S1414 & 1.4 & 1.4 & 2.8& 3.032 & 1.0 & 1.0 &Shen&0.06&3.6&2.64&0.91&2.20&4.6\\
S138142 & 1.38 & 1.42 &2.8& 3.032 & 0.97 & 0.97&Shen&0.06&3.6&2.64&0.89&2.30&5.0\\
S135145 & 1.35 & 1.45 &2.8& 3.034 & 0.93 & 0.93&Shen&0.09&5.9&2.62&0.88&2.35&5.3\\
S1315 & 1.3 & 1.5 &2.8& 3.037 & 0.87 &0.86&Shen&0.15&10&2.57&0.85&2.4&5.7\\
S1216 & 1.2 & 1.6 &2.8& 3.039 & 0.75 &0.73&Shen&0.23&17&2.50&0.77&2.6&6.9\\
S1515 & 1.5 & 1.5 &3.0& 3.274 & 1.0 &1.0&Shen& 0.05&3.4&2.82&0.90&2.4&5.3\\
S1416 & 1.4 & 1.6 &3.0& 3.274 & 0.88 &0.86&Shen&0.17&12&2.73&0.86&2.5&6.2\\
S1317 & 1.3 & 1.7 &3.0& 3.279  & 0.76&0.75&Shen&0.23&17&2.69&0.79&2.7&7.0\\
S119181& 1.19 & 1.81 & 3.0 & 3.289 & 0.66 & 0.63 &Shen&0.24&20&2.67&0.73&2.85&7.9\\
S107193& 1.07 & 1.93 & 3.0 & 3.306 & 0.55 & 0.52 &Shen&0.26&24&2.67&0.64&3.1&9.5\\
S1313 & 1.3 & 1.3 &2.6& 2.800 & 1.0 &1.0&Shen&0.08&4.7&2.44&0.91&2.0&4.3\\
S1214 & 1.2 & 1.4 &2.6& 2.799 & 0.86 &0.85&Shen&0.20&13&2.33&0.84&2.2&5.3\\
S1115 & 1.1 & 1.5 &2.6& 2.807 & 0.73 &0.71&Shen&0.23&17&2.31&0.77&2.35&6.2\\
\hline
C1216 & 1.2 & 1.6 &2.8& 3.039 & 0.75 &0.73&Shen\_c&0.21&21&2.39&0.75&2.5&6.7\\
C1315 & 1.3 & 1.5 &2.8& 3.037 & 0.87 &0.86&Shen\_c&0.14&15&2.47&0.84&2.35&5.8\\

\end{tabular}
\end{center}
\end{table*}

\section{Results of Merger Models}

\subsection{Dynamics of Merging}
 
The evolution of a NS binary during the adiabatic inspiral is
driven by gravitational radiation reaction. Once the binary
becomes unstable to tidal forces, the hydrodynamic evolution sets in,
leading to the tidal stretching of one or of both companions and to
the formation of a central hypermassive NS surrounded by a thick,
neutron-rich accretion torus. Depending on its mass and its
rotation rate and the EoS, the
hypermassive NS will collapse immediately or after a (short?) delay to a
rapidly rotating BH (e.g. \citealt{morrison2004}).

The dynamics and post-merging structure largely depend on the mass ratio $q$
of the binary. In asymmetric systems with $q$ significantly smaller than $1$,
the less massive but slightly larger star is tidally disrupted and deformed into an
elongated primary spiral arm, which is mostly accreted onto the more massive
companion. Its tail, however, contributes a major fraction to the
subsequently forming thick disc/torus around a highly deformed and
oscillating central remnant (see Fig.~\ref{fig:particles}, panels (a) and (b)). 
In systems with $q\simeq1$ and nearly equally sized stars, both stars are 
tidally stretched but not disrupted and directly plunge together into a
deformed merger remnant (see Fig.~\ref{fig:particles2}). 
In all models, excited radial and non-radial
oscillations of the compact remnant periodically lead to mass shedding
off the surface and to ejection of material into secondary tidal tails during the subsequent evolution.
As illustrated in Figs.~\ref{fig:particles} and \ref{fig:particles2},
this effect takes place both in
symmetric and asymmetric cases, but is stronger in the latter
because of the larger triaxial deformation of the post-merger object.

\begin{figure}
\begin{center}
   \includegraphics[width=7.5cm]{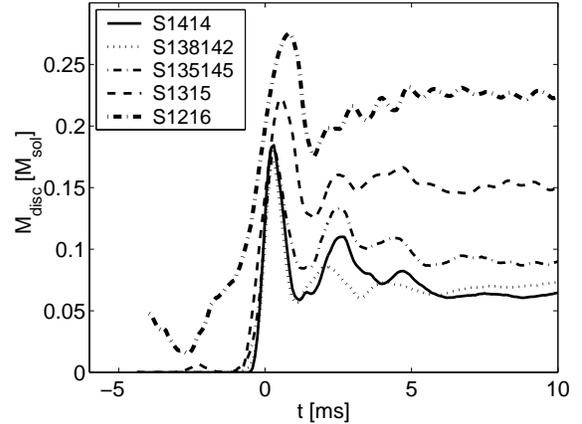}
   \caption{Disc formation in models with non-zero temperature
EoS and $M_{\mathrm{sum}} = 2.8\,M_\odot$.
Plotted is the disc mass as inferred from our disc mass
criterion (see text). The final disc mass as given in
Table~\ref{tab:inittable} is read off at about 6ms. Clearly visible
is the rapid rise at $t\simeq 0$, which depends strongly on the   
mass ratio. On the other hand, the further evolution after merging
is similar in all cases.}
   \label{fig:discmasses_massratio}
\end{center}
\end{figure}

\begin{figure}
\begin{center}
   \includegraphics[width=7.5cm]{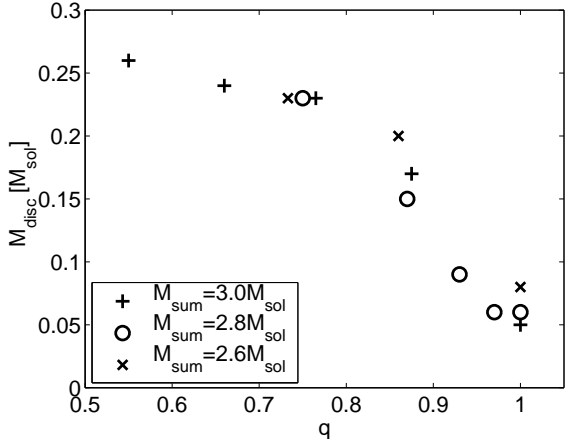}
   \caption{Disc masses versus mass ratio $q$ for all non-zero temperature
   runs.}
   \label{fig:discmasses_summary}
\end{center}
\end{figure}
 
\begin{figure}
\begin{center}
   \includegraphics[width=7.5cm]{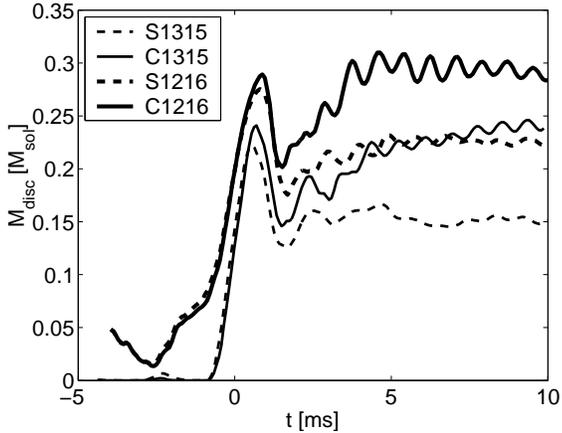}
   \caption{Same as in Fig. \ref{fig:discmasses_massratio}, but comparing
the differences between $T=0$ EoS and non-zero temperature EoS for
two different mass ratios $q$.}
   \label{fig:discmasses_hot_vs_cold}
\end{center}
\end{figure}

\begin{figure}
\begin{center}
   \includegraphics[width=7.5cm]{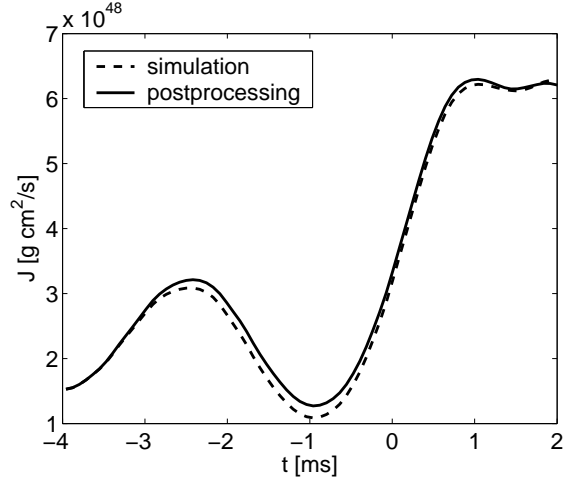}
   \caption{Total angular momentum of a blob of matter in the primary
spiral arm of Model C1216. The dashed line represents the results   
of the simulation, the solid line those of a postprocessing analysis.}
   \label{fig:postprocessing}
\end{center}
\end{figure}

\subsection{Disc Formation}
\label{sect:disc}

If and when the compact post-merger object collapses to a BH (the secular evolution driven by GW radiation reaction, viscosity and MHD cannot be followed by 3D simulations), pressure support from the central remnant to the disc will vanish and matter will be prevented from infall mostly by rotational support. As a consequence, we distinguish future disc matter from the central remnant by demanding the specific angular momentum $j$ to be larger than the one associated with the innermost stable circular orbit (ISCO) of a Kerr BH with the gravitational mass and the spin parameter of the central remnant.\\

The ISCO can be analytically
determined in the case of the Boyer-Lindquist Kerr metric
\citep{bardeen1972}. In our case we shall use an approximative pseudo-Kerr
metric (see, e.g., \citealt{grandclement2002}) which is both isotropic and
conformally flat, consistent with the coordinates we used for our numerical
simulations. The ISCO can then be found among 
all circular orbits in the orbital plane by minimizing the specific angular
momentum along the radial coordinate.

Using this criterion, we are now able to identify the disc matter and disc mass for the various models considered. Their very different merger dynamics is reflected in
the different evolution of the disc mass (see Fig.~\ref{fig:discmasses_massratio}). The rapid rise in the asymmetric models around $t\simeq0$ms can be associated with the
development of the primary spiral arm. After this dynamical phase
shortly after merging, a nearly stationary torus forms and the disc mass 
settles down to a slowly changing value, which is plotted versus mass ratio $q$ for
all models in Fig.~\ref{fig:discmasses_summary}. We find a rapid rise
of the disc mass when $q$ drops below unity, and a flattening for
$q$-values below $\sim 0.8$. The disc mass increases from about 
$0.05M_{\odot}$ at $q=1$ to about $0.26M_{\odot}$ at $q=0.55$.


We checked the consistency of our disc mass determination by
comparing the obtained masses with the amount of matter residing
outside the equatorial radius of the remnant $r_{\mathrm{rem}}$. We determine
$r_{\mathrm{rem}}$ in a radial density profile by locating the point between
the steep density gradient at the remnant surface and the slower fall-off in
the disc. In all cases, we found $M_{\mathrm{gas}}(r>r_{\mathrm{rem}})>M_{\mathrm{gas}}(j>j_{\mathrm{ISCO}})\equiv M_{0,\mathrm{disc}}$, suggesting that gas pressure still plays an important role in the quasi-stationary phase. Our disc masses are therefore lower limits, since pressure support will not completely vanish even after BH formation.  

\begin{figure*}
\begin{center}
  \begin{minipage}[t]{0.32\linewidth}
   \includegraphics[width=5.5cm]{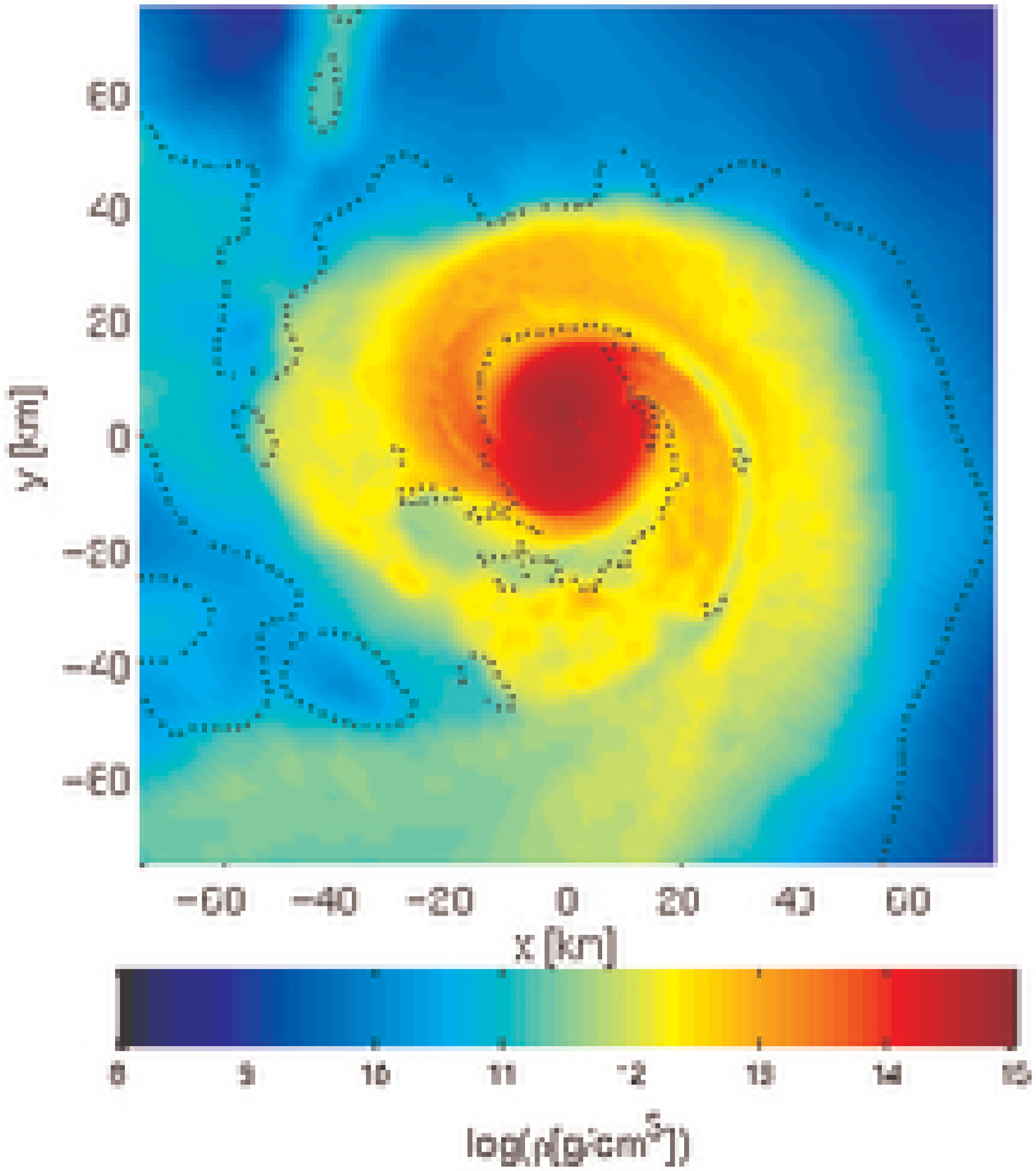} 
\end{minipage}
  \begin{minipage}[t]{0.32\linewidth}
   \includegraphics[width=5.5cm]{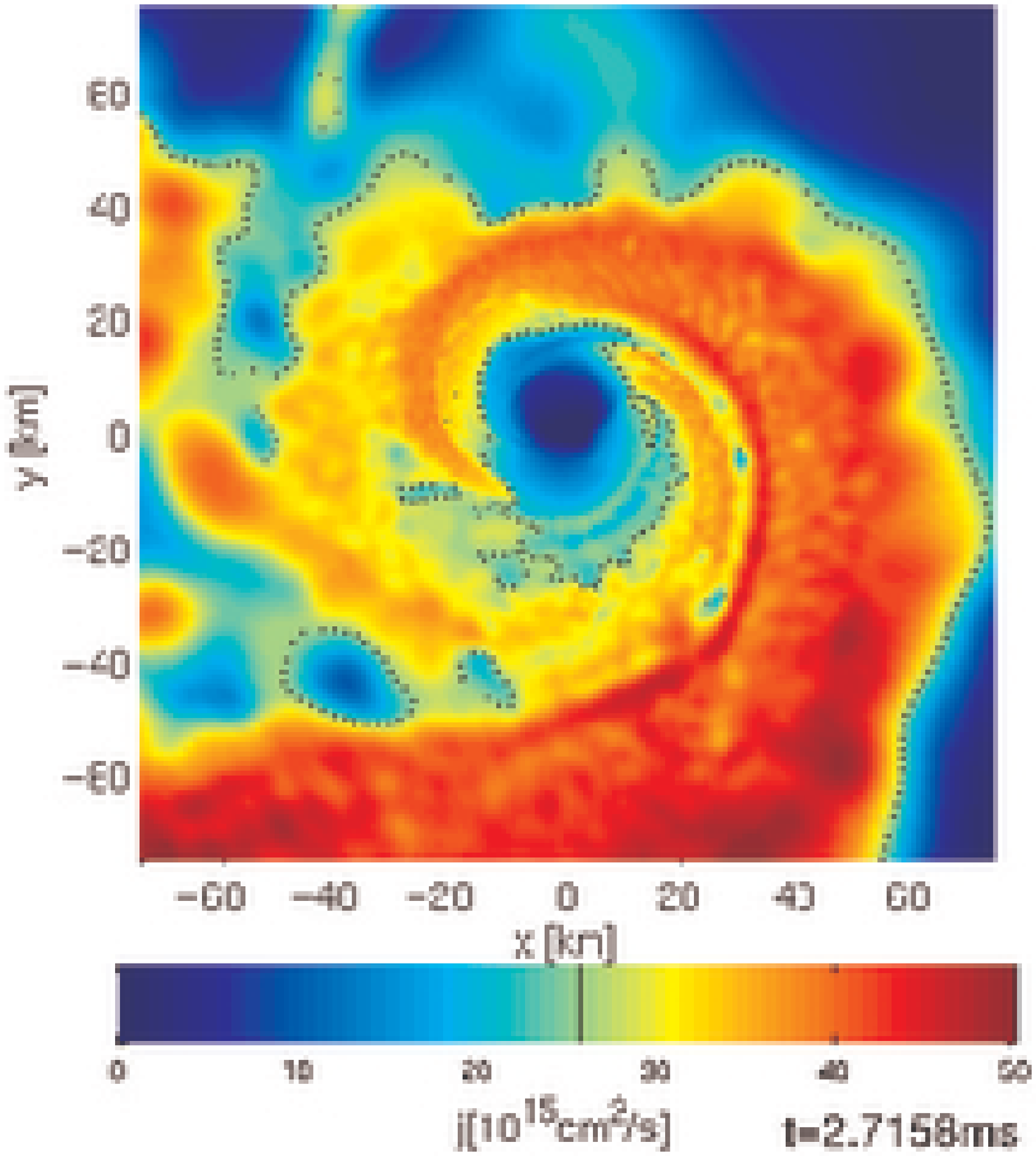} 
\end{minipage}
  \begin{minipage}[t]{0.32\linewidth}
   \includegraphics[width=5.5cm]{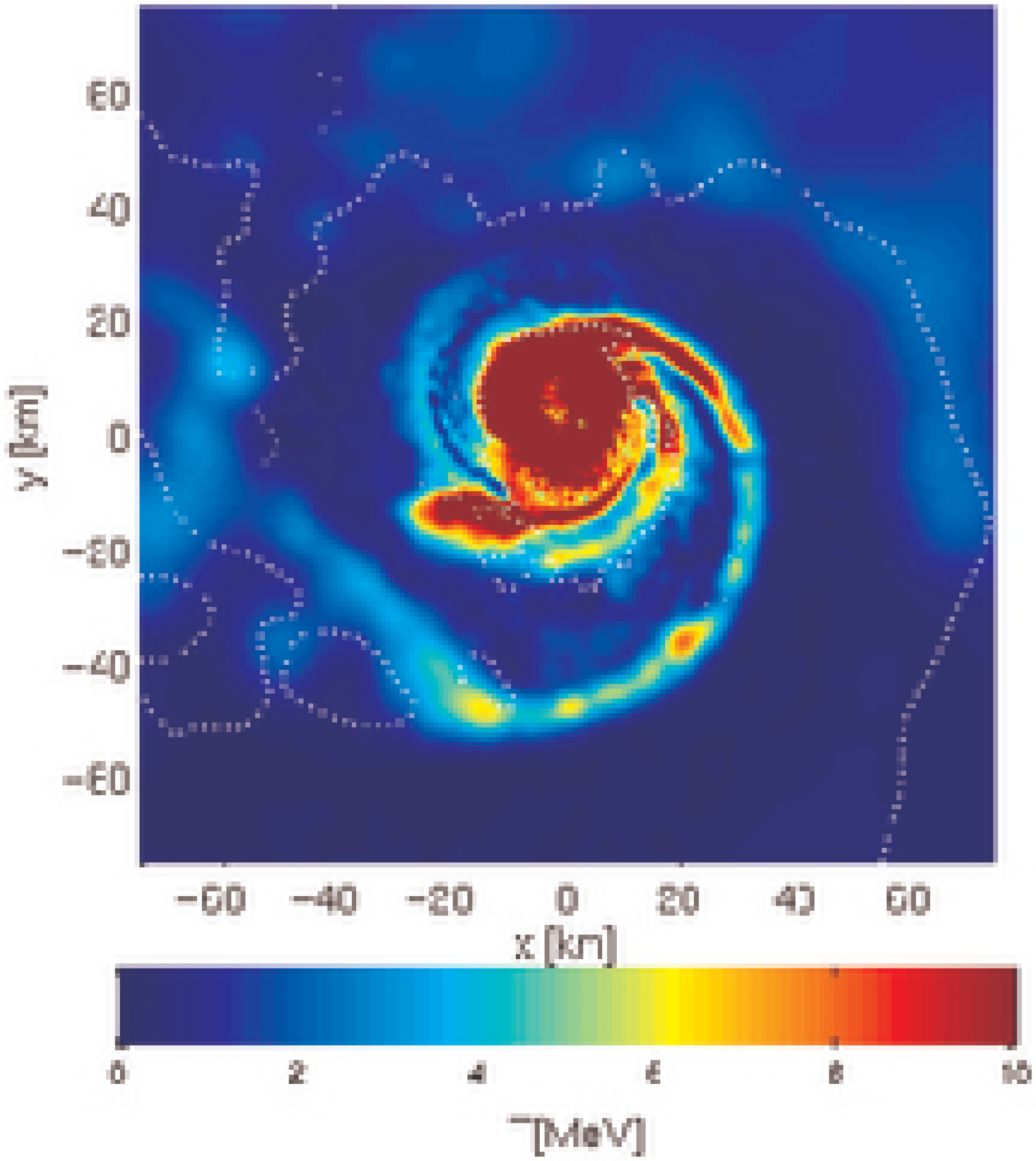} 
\end{minipage}
  \begin{minipage}[t]{0.32\linewidth}
   \includegraphics[width=5.5cm]{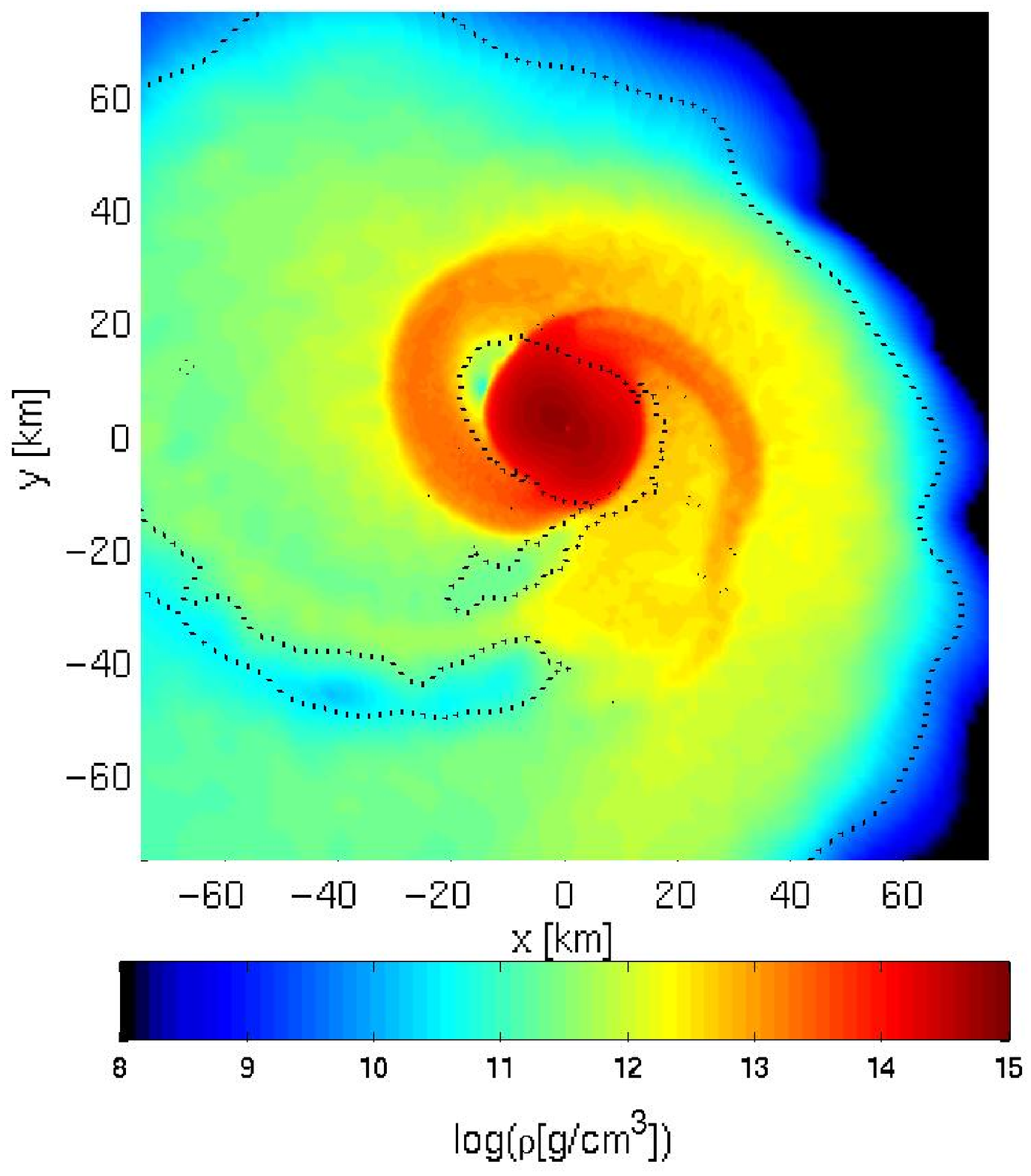} 
\end{minipage}
  \begin{minipage}[t]{0.32\linewidth}
   \includegraphics[width=5.5cm]{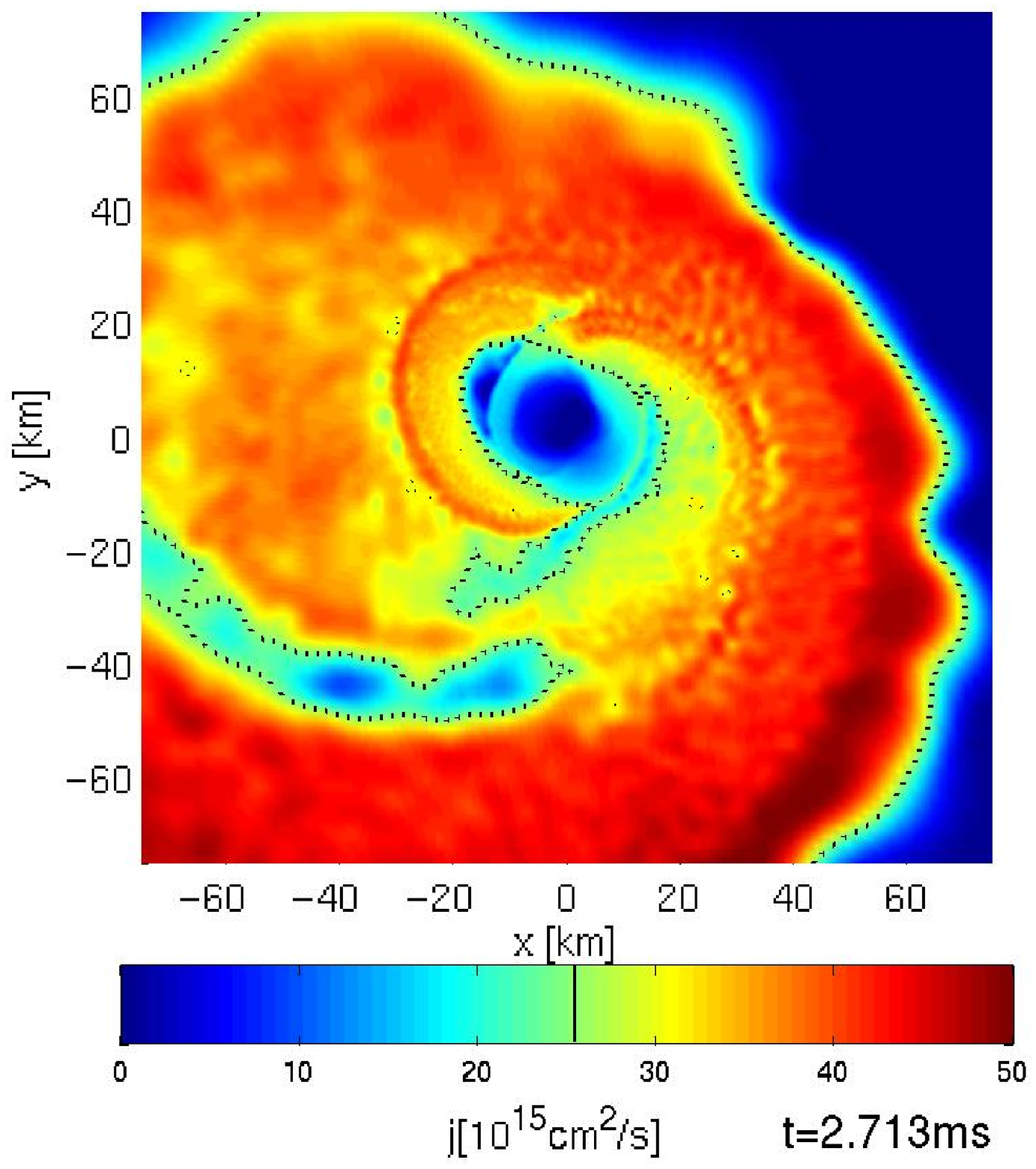} 
\end{minipage}
\begin{minipage}[t]{0.32\linewidth}
	\vspace{-6.1cm}
	\hspace{0.5cm}
	\begin{minipage}[h]{0.95\linewidth}
	   \caption{ 
	    Density $\rho$ (left), specific angular momentum $j$
 	  (middle) and temperature distribution $T$ (right) in the orbital plane
 	  for the $T\neq0$ Model S1216 (top) and the $T=0$ Model C1216
 	  (bottom, without temperature).
 	  The black dotted line in all
 	  panels and the black solid line in the color bars for $j$ correspond to the location where the specific angular momentum is equal to the value of
 	  $j_{\mathrm{ISCO}}$ at the given time (see text). Note that
	the color bar for the upper right panel is limited to 10~MeV
	such that the temperature distribution in the disk becomes
	visible. The maximum temperature in the merger remnant rises
	up to $\sim$30~MeV at the given time.
   	  \label{fig:morph} }	
	\end{minipage}
\end{minipage}
\end{center}
\end{figure*}

\begin{figure*}
 \begin{center}
  \begin{minipage}[t]{0.47\linewidth}
        (a)
   \includegraphics[width=6.7cm]{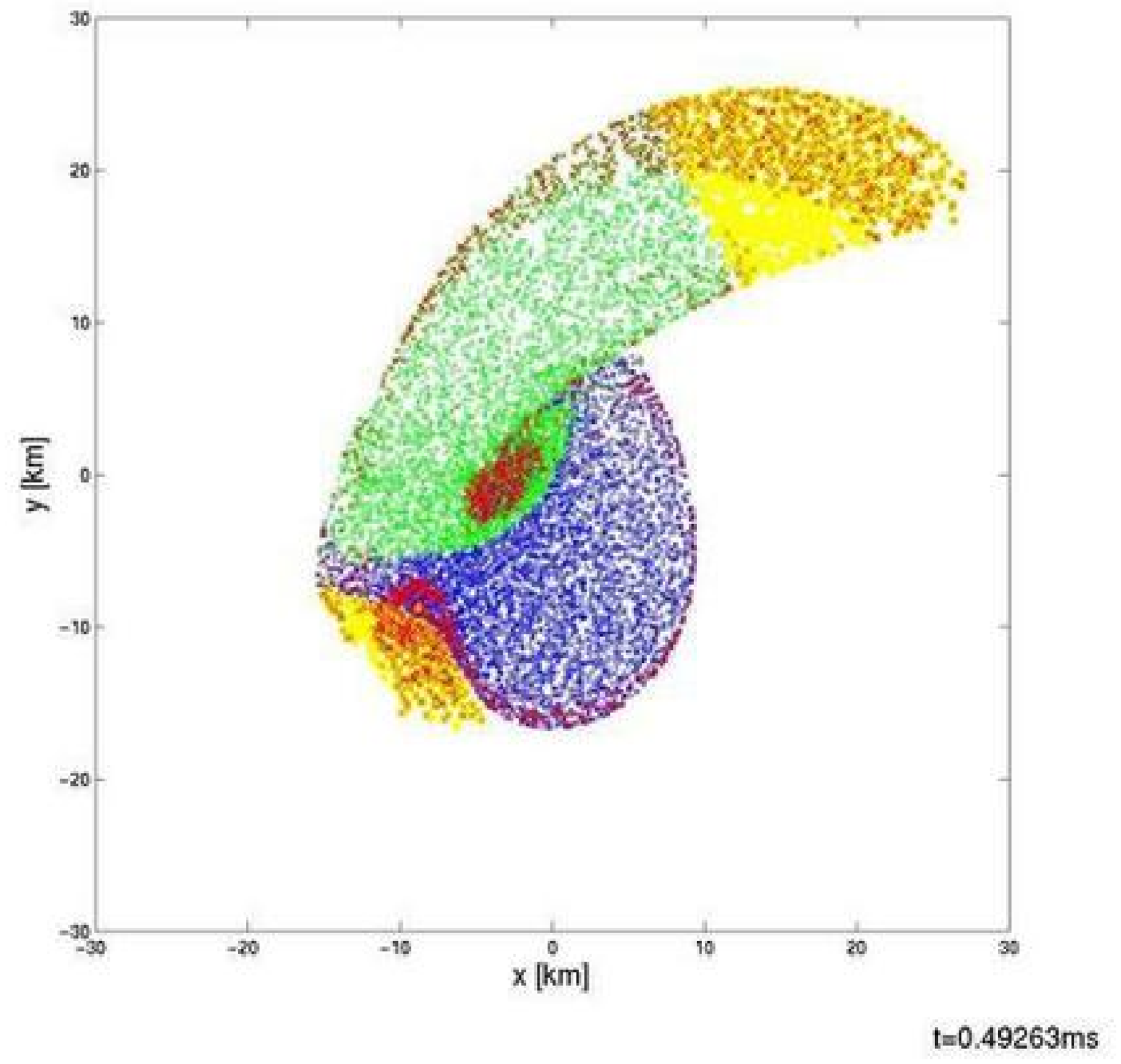} 
\end{minipage}
\begin{minipage}[t]{0.47\linewidth}
        (b)
   \includegraphics[width=6.7cm]{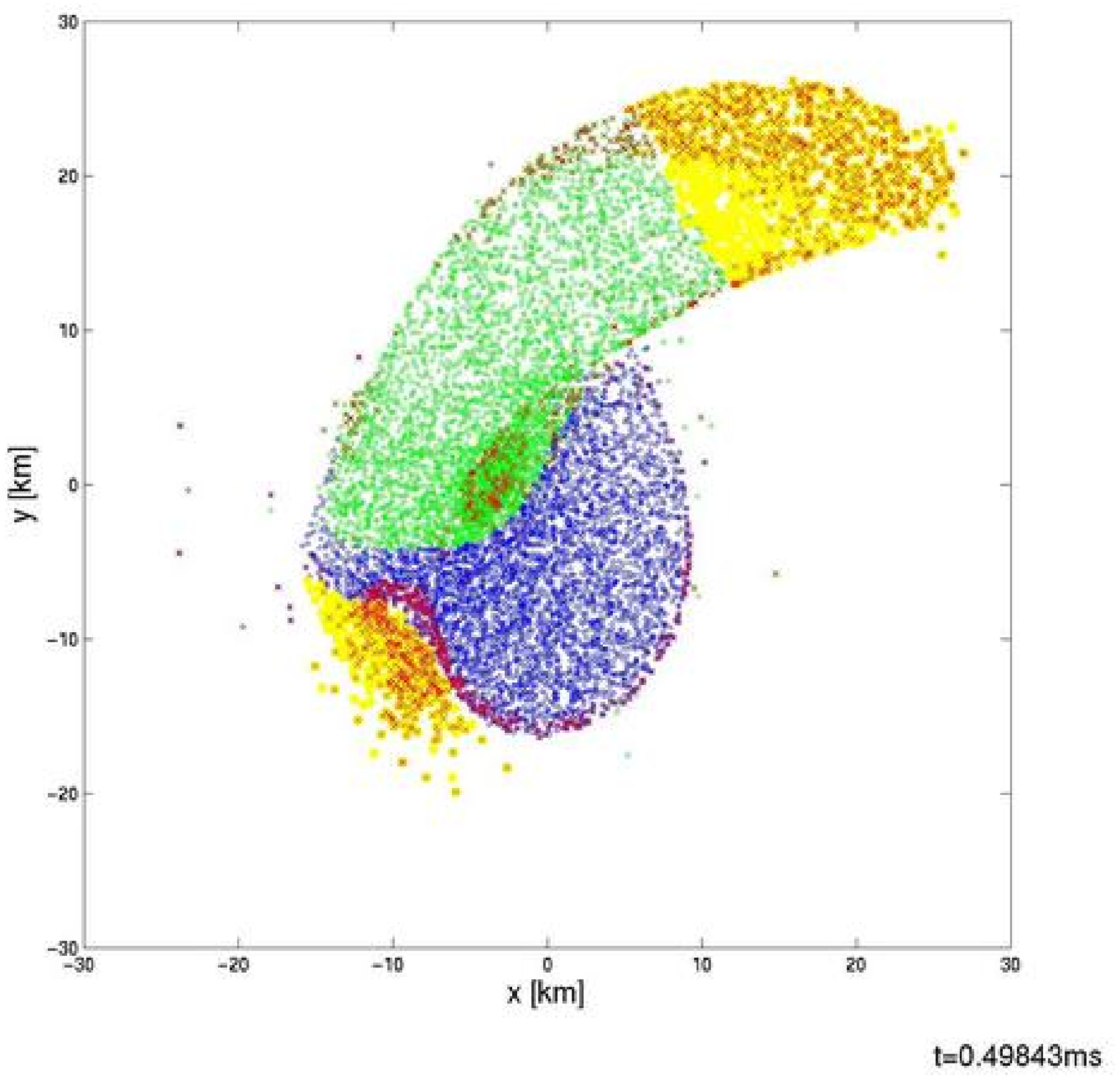}
\end{minipage}
\begin{minipage}[t]{0.47\linewidth}
        (c)
   \includegraphics[width=6.7cm]{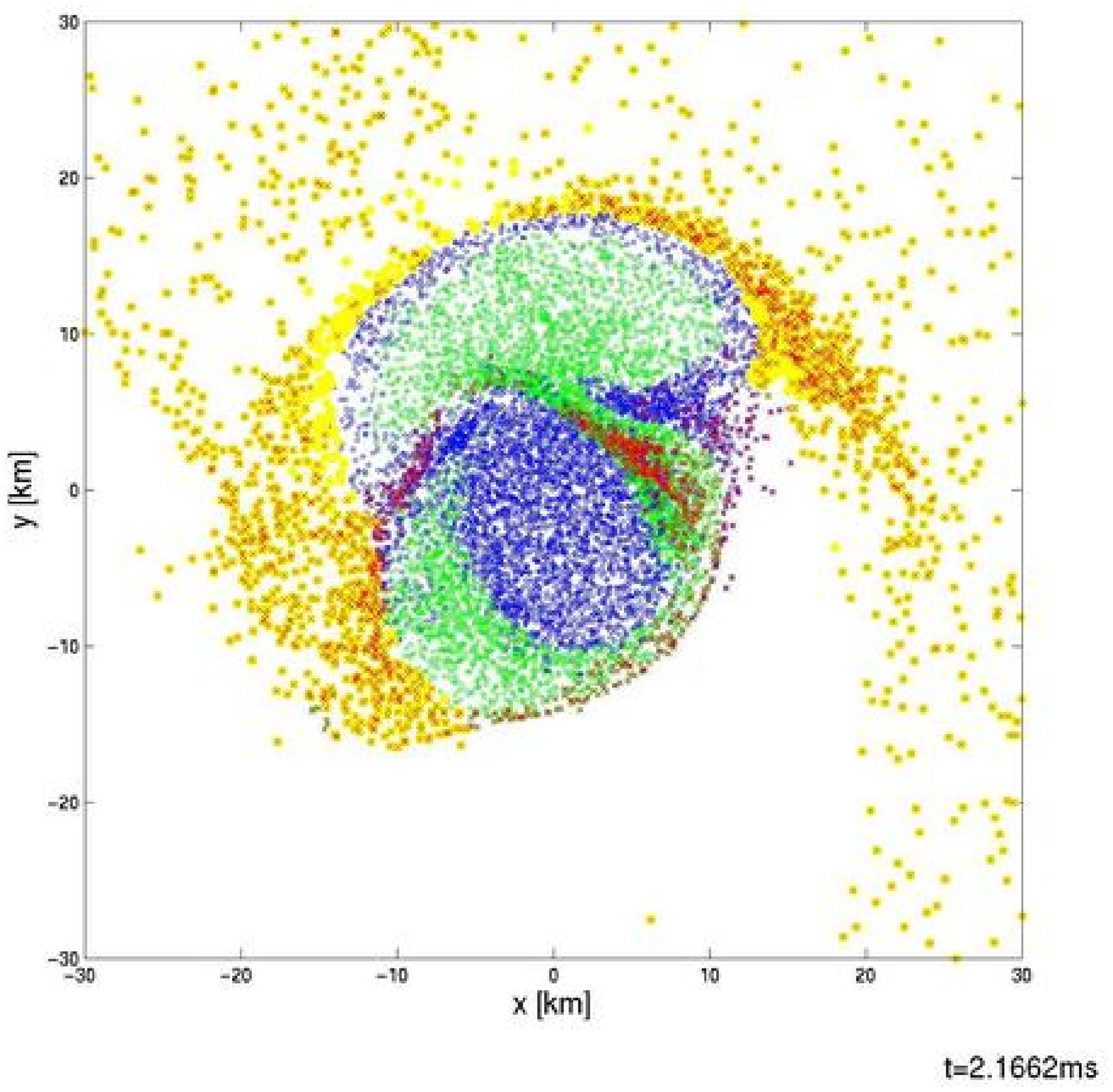}
\end{minipage}
\begin{minipage}[t]{0.47\linewidth}
        (d)
   \includegraphics[width=6.7cm]{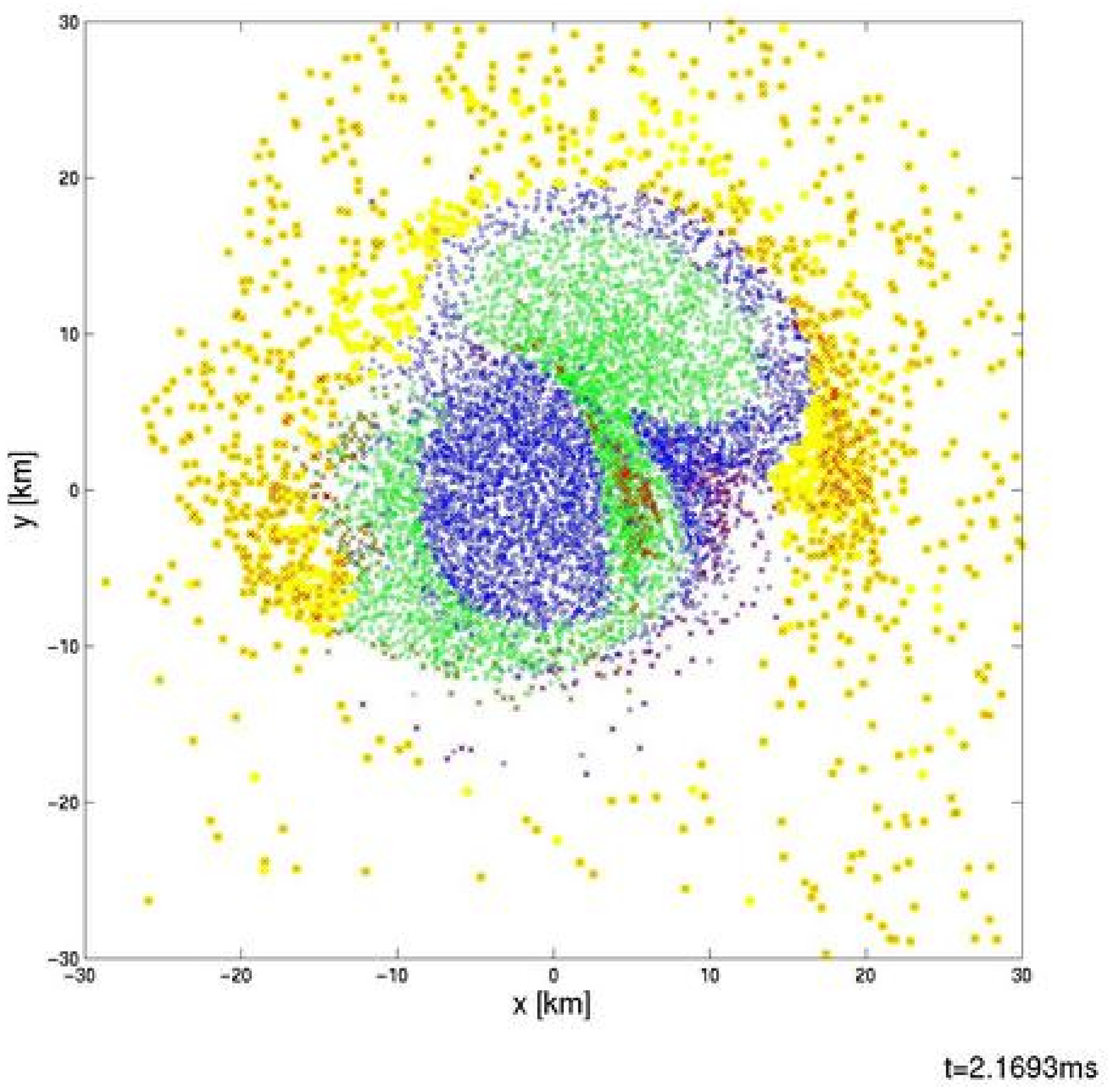}   
\end{minipage}
\begin{minipage}[t]{0.47\linewidth}
        (e)
   \includegraphics[width=6.7cm]{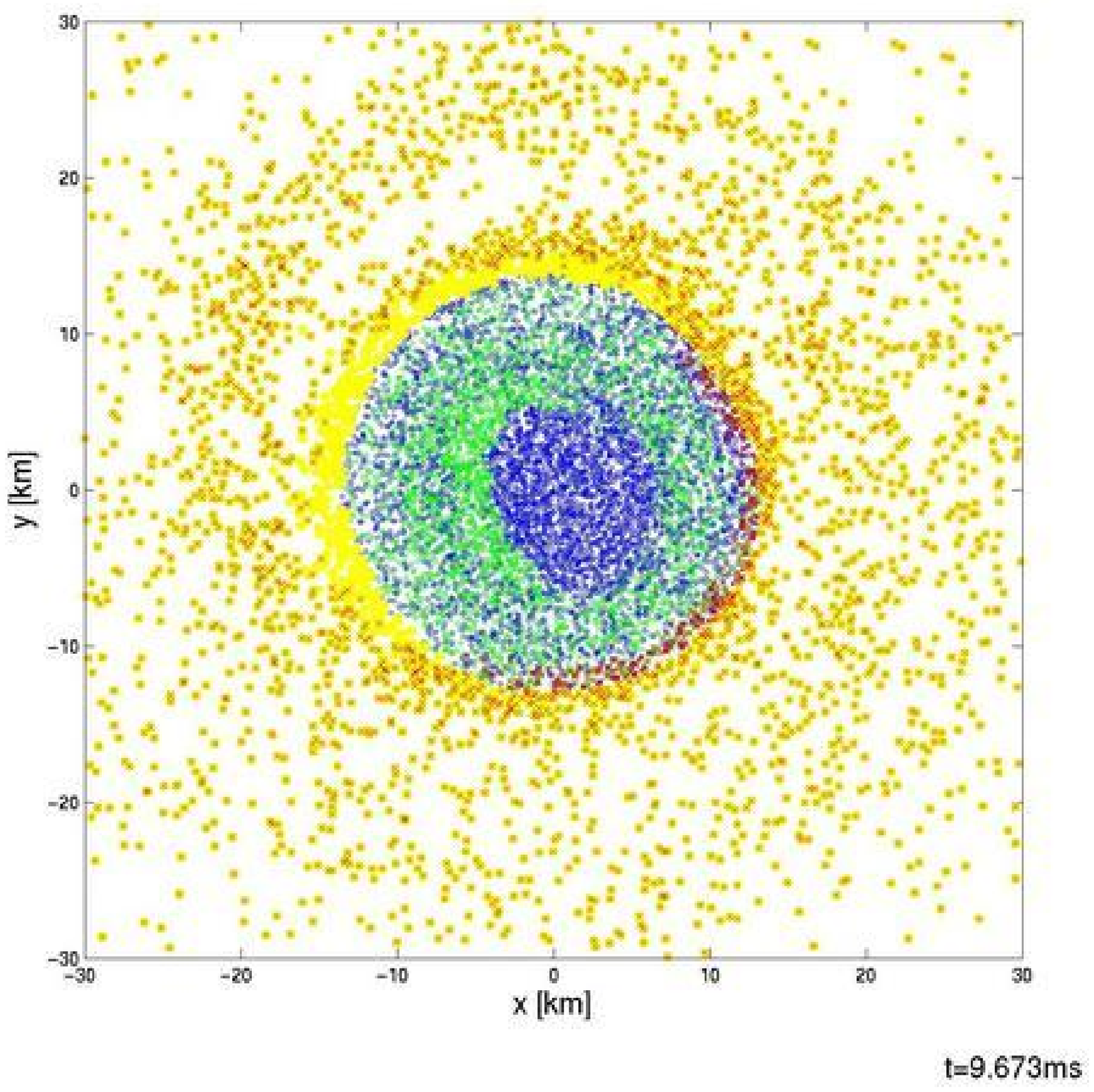}
\end{minipage}
\begin{minipage}[t]{0.47\linewidth}  
        (f)
   \includegraphics[width=6.7cm]{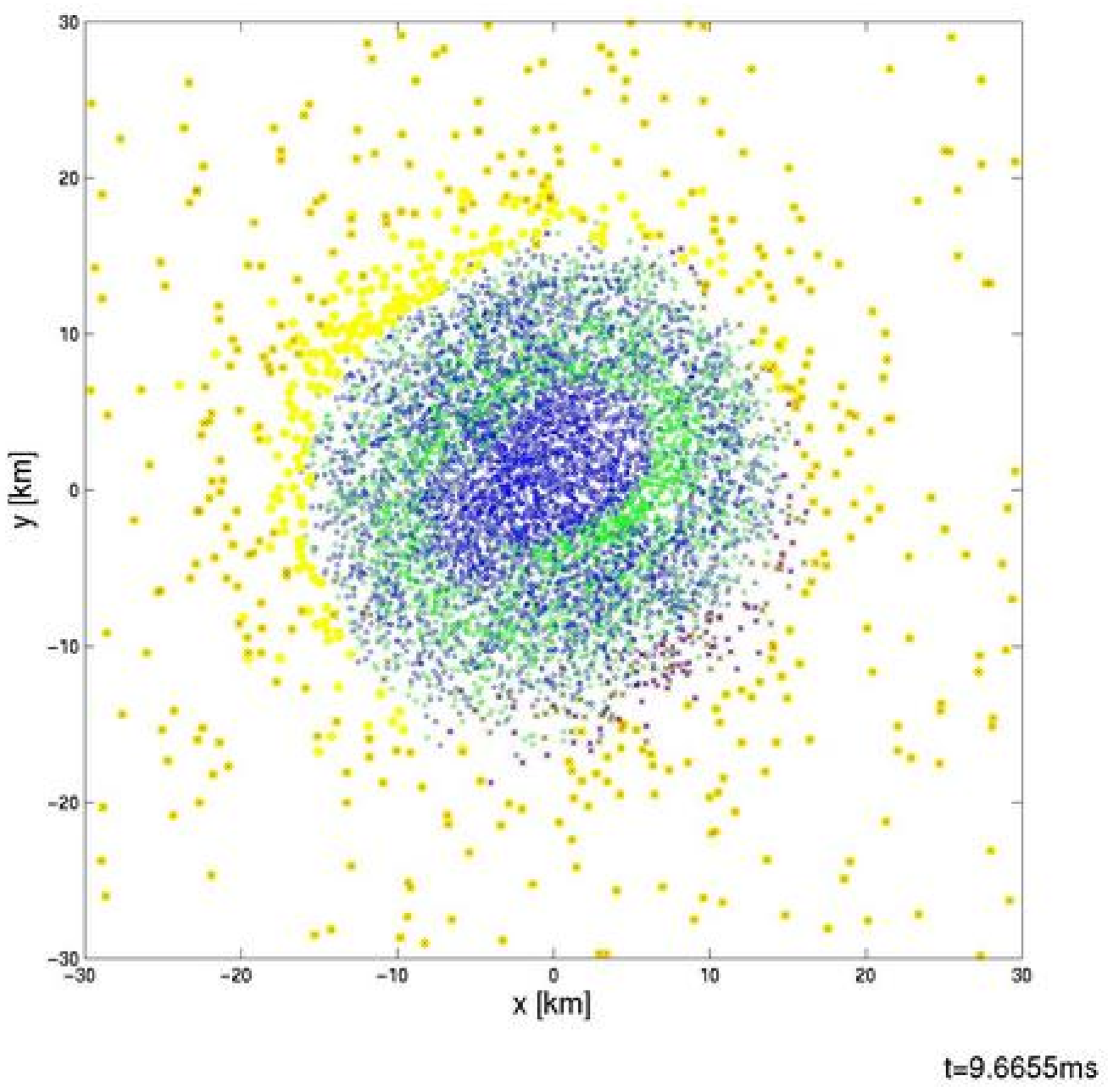}
\end{minipage}
  \caption{Characteristic evolutionary phases of Model C1216
(left panels) and Model S1216 (right panels). Panels (a) and (b)
show the merging phase with a primary spiral arm forming. Plotted is
every 10th SPH particle around the equatorial plane. The matter of the two
stars is represented by green and blue particles, respectively. Particles
that end up in the disc at the end of the simulation are marked in red
and the ones that currently fulfill the disc criterion (see
Sect. \ref{sect:disc}) are plotted in yellow.
Panels (c) and (d) display the situation $\sim2\,$ms after merging, when a
secondary spiral arm appears.
Panels (e) and (f) show the time when a quasi-stationary torus has
formed. Note that the particle density is not strictly proportional
to the density of the fluid because the particles near
the initial stellar boundaries are attributed a smaller particle
mass. These particles preferentially end up in the torus so that
the particle density there appears enlarged.}  
 \label{fig:particles}
\vspace{-0.5cm}
\end{center}
\end{figure*}

\begin{figure*}
 \begin{center}
  \begin{minipage}[t]{0.47\linewidth}
        (a)
   \includegraphics[width=6.7cm]{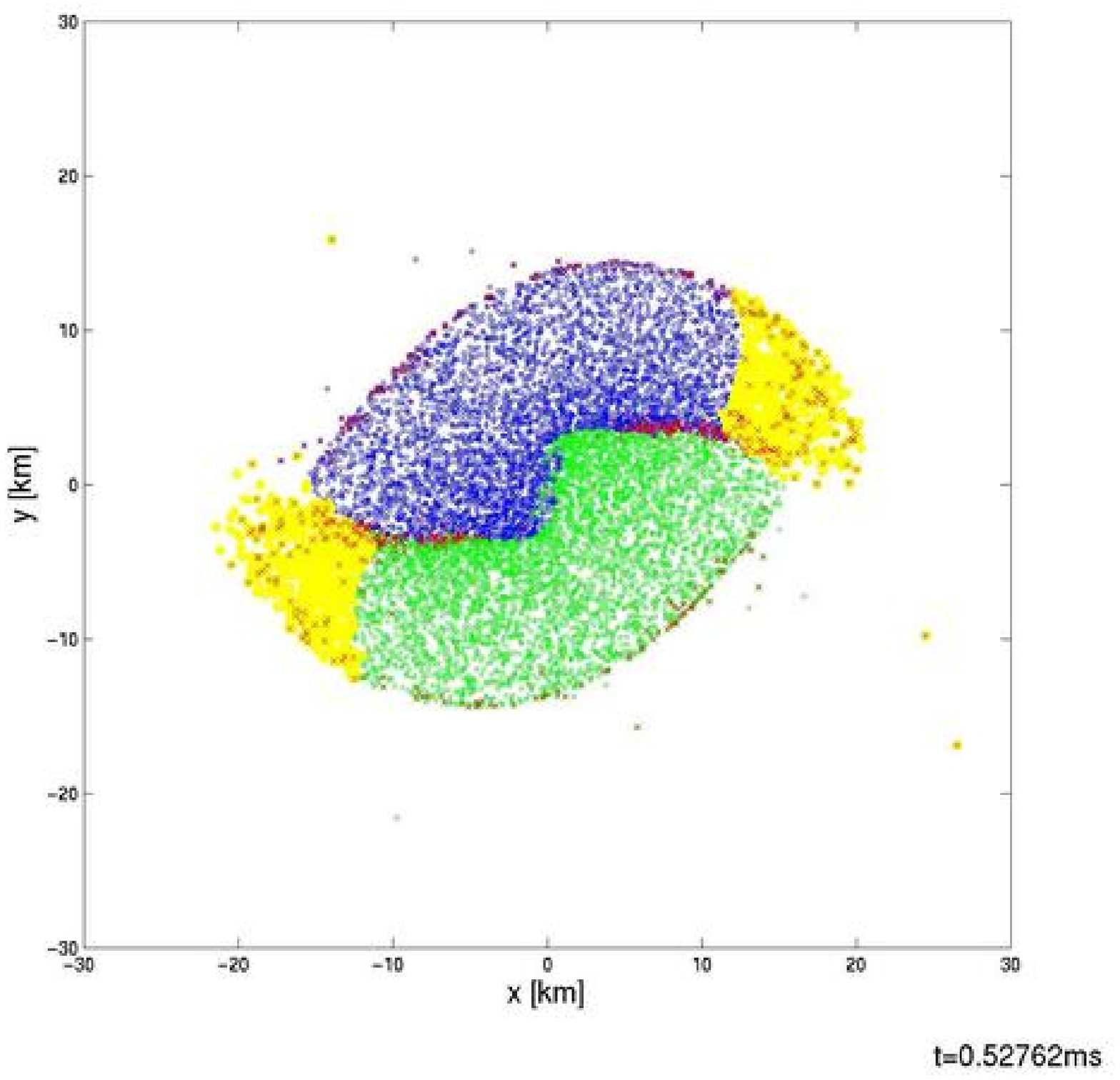}
\end{minipage}
\begin{minipage}[t]{0.47\linewidth}
        (b)
   \includegraphics[width=6.7cm]{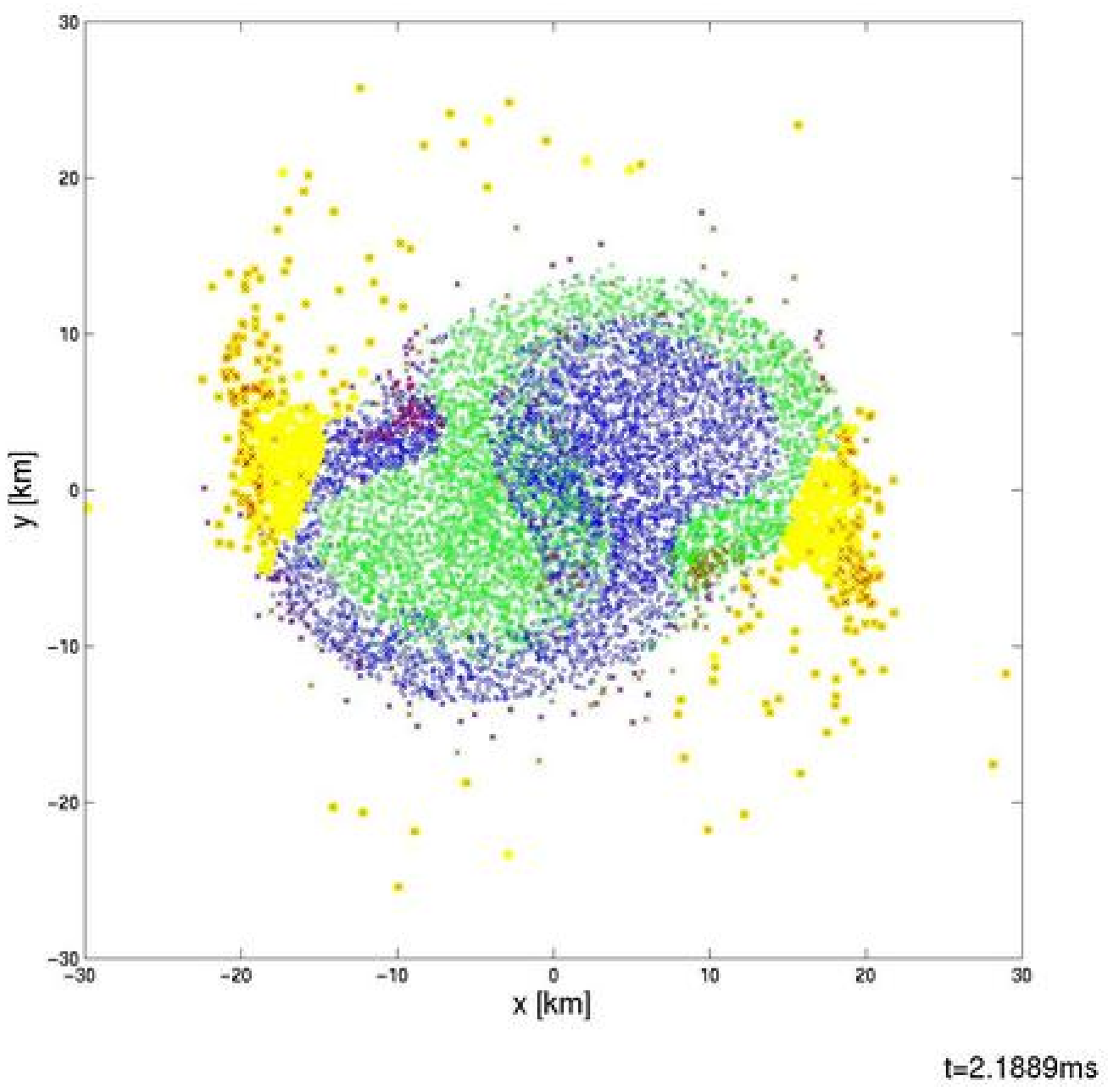}
\end{minipage}
  \caption{
 Same as panels (a) \& (c) and (b) \& (d), respectively, of
 Fig.~\ref{fig:particles}, but for Model S1414.
 The primary spiral arm is absent in the symmetric binary case
 and the post-merging (secondary) arms are smaller than in the
 asymmetric models}

.
 \label{fig:particles2}
\vspace{-0.5cm}
\end{center}
\end{figure*}

In Fig.~\ref{fig:particles} we have color-coded for a $T=0$ and 
a $T\neq 0$ model in red the material that belongs to the future disc, whereas the
material that currently fulfills the disc criterion is shown in yellow. By definition, these two attributes coincide at the end of the simulation. We see that the future disc matter originates mainly from
the spiral arm tips and from the stellar surface. This indicates that
the formation of a post-merger disc is linked to the presence of
spiral arms. Indeed, the yellow tips of these tidal tails in
Fig.~\ref{fig:particles} and also the enhanced
specific angular momentum in the outer edges of the tidal tails in
Fig.~\ref{fig:morph} suggest angular momentum transfer 
from the central remnant to the spiral arm tips via tidal torques. 
This hypothesis is confirmed by a postprocessing analysis of Model C1216 
where we integrated the torque exerted by gravitational and pressure 
forces from the compact deformed remnant on the yellow mass elements 
in the primary spiral arm as a function of time. The result for the 
total angular momentum evolution agrees very well with the simulation 
result (Fig.~\ref{fig:postprocessing}). 

An interesting effect can also be seen by comparing models with
$T=0$ and $T\neq0$ EoSs (Figs.~\ref{fig:discmasses_hot_vs_cold},
\ref{fig:morph}, \ref{fig:particles}). 
In Fig. \ref{fig:discmasses_hot_vs_cold} we show the
disc mass evolution of the cold Models C1216 and C1315 together with
their non-zero temperature counterparts S1216 and S1315. Obviously,
the first increase in disc mass, which can be associated with the cold
primary spiral arm, is unaffected by temperature
effects. Thermal pressure effects during the following evolution, however,
reduce the further increase in disc mass in the non-zero temperature case,
whereas in the cold case the disc mass increases by an additional
$\sim 0.07M_\odot$. We determine two possible reasons for this difference. 
On the one hand, the core of the cold merger remnant contracts more 
strongly due to the absence of thermal pressure. Therefore, more 
gravitational binding energy is converted into rotational energy and 
the rotation rates become higher, supporting the formation of spiral 
arms. On the other hand, the shock-heated matter from the collision 
region between the two NSs forms a halo of hot matter which engulfs 
the compact remnant-torus system in the $T\neq0$ case 
(Figs.~\ref{fig:morph}, \ref{fig:particles}). This halo may 
damp the oscillations of the remnant and the development of large 
deviations from axisymmetry.

\section{Constraints from Observed GRBs}

In this section we shall discuss how the properties of the 
GRB engine may be constrained from the recent observations
of four short-duration GRBs. To this end we shall make the 
assumption that 
the bursts originated from ultrarelativistic jets whose
acceleration was mainly driven by thermal energy, provided by the 
annihilation of neutrino-antineutrino ($\nu\bar\nu$) pairs
in the vicinity of post-merger BH-torus systems (see, e.g.,
\citealt{ruffert1999}, \citealt{setiawan2004}). Although we will
refer to this scenario in our discussion because numerical models
provide at least order of magnitude information of involved 
parameters, our arguments are more general and apply basically
also to models that consider MHD-powered GRB jets.

\subsection{BH-Torus System Parameters}

Linking the measured apparent (i.e., isotropic-equivalent)
energy $E_{\gamma,\mathrm{iso}}$ of a short GRB to the energy output
from the central engine and the mass $M_{\mathrm{acc}}$ accreted
by the BH (during the phase when the neutrino emission from the
accretion torus is sufficiently powerful to drive the jets) involves
a chain of efficiency parameters corresponding to the different
steps of physical processes between the energy
release near the BH and the gamma-ray emission
at $\sim\,$10$^{14}\,$cm$\,$:
\begin{equation}
E_{\gamma,\mathrm{iso}}\,=\, f_1\,f_2\,f_3\,f_\Omega^{-1}\,f_4\,
M_{\mathrm{acc}}c^2\ .
\label{eq:torusmass}
\end{equation}
Here $f_1$ denotes the efficiency at which accreted rest mass energy
can be converted to neutrino emission, $f_2$ is the conversion efficiency
of neutrinos and antineutrinos by annihilation to $e^\pm$ pairs, $f_3$
is the fraction of the $e^\pm$-photon fireball energy which drives the
ultrarelativistic outflow with Lorentz factors $\Gamma > 100$ as
required by GRBs, 
$f_\Omega = 2\Omega_{\mathrm{jet}}/(4\pi) = 1 - \cos\theta_{\mathrm{jet}}$
denotes the jet collimation factor defined as 
the fraction of the sky covered by the two polar jets 
(with semi-opening angles $\theta_{\mathrm{jet}}$ and solid angles
$\Omega_{\mathrm{jet}}$),
and $f_4$ is the fraction of the energy of
ultrarelativistic jet matter which can be emitted in gamma rays in course
of dissipative processes that occur in shocks when optically thin
conditions are reached. 

These parameters are constrained to some
degree by numerical and analytic work, but their values are still
rather uncertain and might vary strongly with time-dependent and 
system-dependent
conditions of the source. ``Typical'' values from merger and
accretion simulations are: $f_1\sim 0.05$ (\citealt{lee2005b}, \citealt{setiawan2004}),
$f_2\sim 0.001\,...\,0.01$ (\citealt{ruffert1999}, 
\citealt{setiawan2004}), $f_3\sim 0.1$, 
$f_\Omega\sim 0.01\,...\,0.05$, and $f_4\la 0.2$ from estimates for 
internal shock models (\citealt{daigne1998}, \citealt{kobayashi2001}, \citealt{Guetta2001}, and references therein).

The cited values for $f_3$ and $f_\Omega$ were determined by 
relativistic hydrodynamical calculations of the formation and
propagation of jets driven by
thermal energy deposition around BH-torus systems
\citep{aloy2005}. These simulations revealed that about 
10--30\% of the $e^\pm$-pair
plasma fireball energy are used for hydrodynamically accelerating the
baryonic matter in the jets to ultrarelativistic velocities.
Typical jet half-opening angles
were found to be between 10$^\mathrm{o}$ and 15$^\mathrm{o}$, corresponding
to $f_\Omega \sim 1.5\times 10^{-2}\,$...$\,3.4\times 10^{-2}$ 
(\citealt{aloy2005}, \citealt{janka2005}).
Two of the four well-localized short GRBs indeed provide hints for
this degree of collimation (see \citealt{fox2005}, 
\citealt{berger2005}) predicted by the theoretical work.

\begin{figure}
   \includegraphics[width=7.5cm]{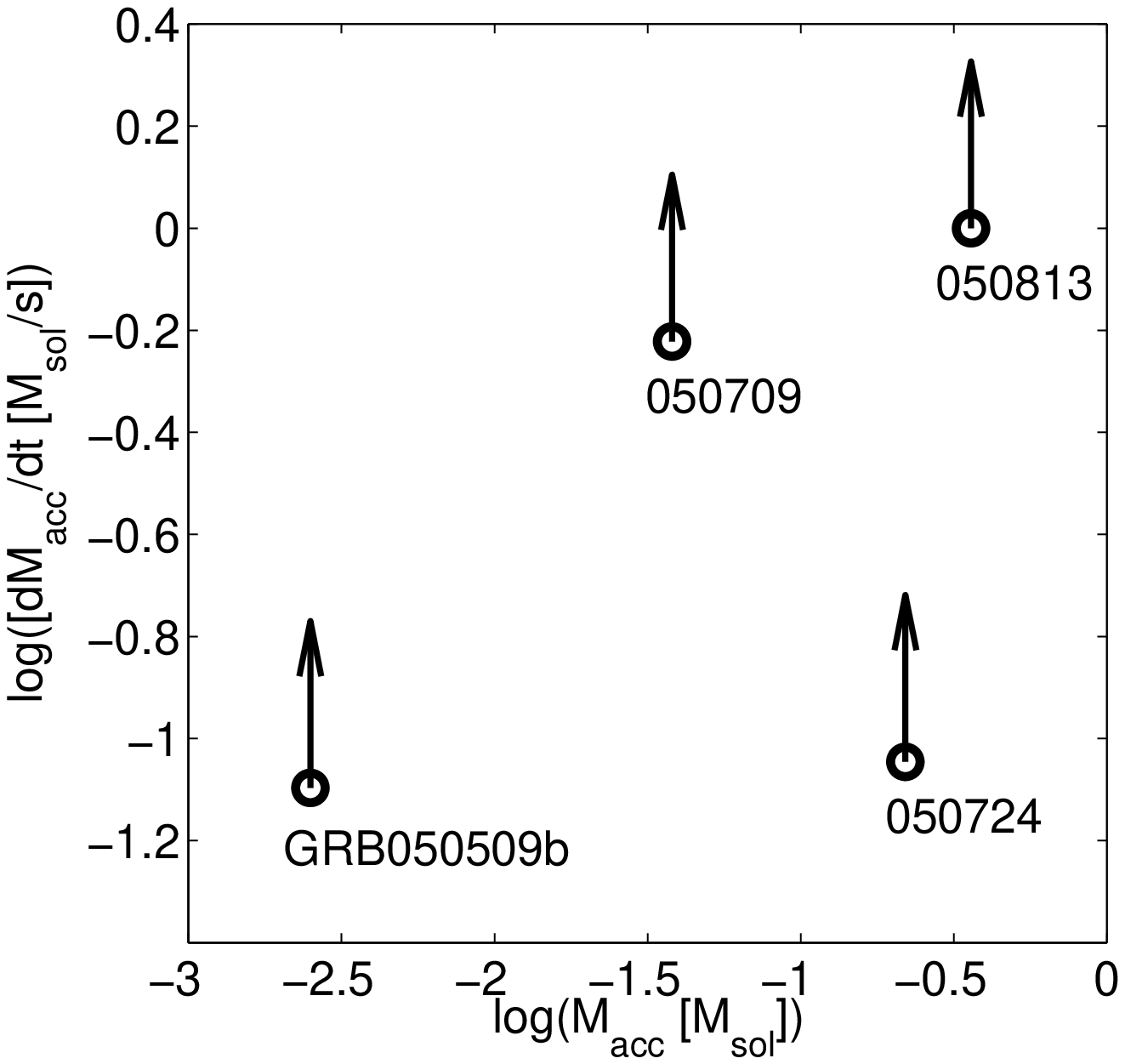}
   \caption{Estimated accreted masses, $M_{\mathrm{acc}}$, and lower
bounds for the average mass accretion rates,
$\dot M_{\mathrm{acc}}$ for the four well-localized short GRBs of
Table~\ref{tab:discmasses}.}
   \label{fig:accmasses}
\end{figure}

Because of the lack of detailed information about how the factors
$f_1$ to $f_4$ and $f_\Omega$ depend on the properties of 
NS+NS binaries and their relic BH-torus systems, we made the
simplifying and bold assumption that the set of parameter values
is the same for all cases, 
$(f_1,\,f_2,\,f_3,\,f_\Omega,\,f_4) = (0.1,\,0.01,\,0.1,\,0.01,\,0.1)$,
and applied Eq.~(\ref{eq:torusmass}) to the four observed short GRBs,
using the measured isotropic-equivalent gamma-energies of
Table~\ref{tab:discmasses}. Doing so, we obtained the estimates 
listed in that table for the accreted mass $M_{\mathrm{acc}}$\footnote{ 
Note that the (initial) torus mass may be larger than the accreted mass 
that we determine from the observed GRB properties. 
A part of the torus may fall into the BH without producing
a release of energy which is strong enough to power ultrarelativistic
GRB-jets. Some part of the torus mass will also be ejected by winds
driven by neutrino energy deposition (see e.g., \citealt{ruffert1997}, 
\citealt{rosswog2003}), viscous heating, or MHD effects 
(see, e.g., \citealt{daigne2002}),
and some part of the torus mass may be lost equatorially due to the
outward transport of angular momentum by viscous shear
in the (magnetized) accretion torus.}.
Assuming further that the duration $t_\gamma$ of the GRB (at the
location of the redshifted
source) is an upper bound for the duration of the accretion of
$M_{\mathrm{acc}}$ by
the BH, $t_{\mathrm{acc}} \la t_\gamma$ (cf.~\citealt{aloy2005}), 
we can deduce lower limits for the average mass accretion rates,
\begin{equation}
\dot M_{\mathrm{acc}}\,\ga\,{M_{\mathrm{acc}}\over t_\gamma}\ .
\label{eq:accrate}
\end{equation}
These are also listed in Table~\ref{tab:discmasses} and 
plotted versus $M_{\mathrm{acc}}$ in Fig.~\ref{fig:accmasses}.
We obtain in all cases values for the torus masses 
in the range of those determined in this work
(Fig.~\ref{fig:discmasses_summary}), 
and values for the mass accretion rates in the ballpark of
theoretical expectations for post-merger BH accretion. Simulations
of the latter find mass accretion rates of typically between
several $0.1\,M_\odot\,$s$^{-1}$ and several $M_\odot\,$s$^{-1}$
(e.g., \citealt{ruffert1999},
\citealt{setiawan2004}, \citealt{lee2002}, \citealt{lee2005b}). 
Of course, the numerical values for $M_{\mathrm{acc}}$ and 
$\dot M_{\mathrm{acc}}$ in Table~\ref{tab:discmasses} are very
uncertain and have to be taken with caution. Shifts by factors of 
a few are easily possible because of probable system-dependent 
variations of the efficiency factors $f_1$ to $f_4$ and of the 
jet collimation factor $f_\Omega$. Nevertheless, 
it is very encouraging that the simple and straightforward 
calculations based on Eqs.~(\ref{eq:torusmass}) and (\ref{eq:accrate})
lead to reasonable numbers in all cases.

GRB~050509b, for example, is particularly interesting because of
its very short intrinsic duration of 
$t_\gamma\sim 33\,$ms and an extraordinarily low isotropic 
energy of a few 10$^{48}\,$erg if it is located at $z = 0.225$
\citep{gehrels2005}. Such a low-energy event
can be explained by the accretion of discs in the lower range
of masses found here for symmetric NS+NS binaries, even when 
rather conservative assumptions are made for the incompletely known 
efficiencies of the various process steps between the energy
release near the BH and the gamma-ray emission.
We find that an accreted mass of
$10^{-3}$--$10^{-2}\,M_{\odot}$ is well sufficient to account for the
energetics of GRB~050509b. Of course, the mass could also be larger
if the energy conversion to ultrarelativistic outflow and GRB 
emission is less efficient than we assumed.

\begin{figure*}
 \begin{center}
  \begin{minipage}[t]{0.47\linewidth}
        (a)
   \includegraphics[width=6.7cm]{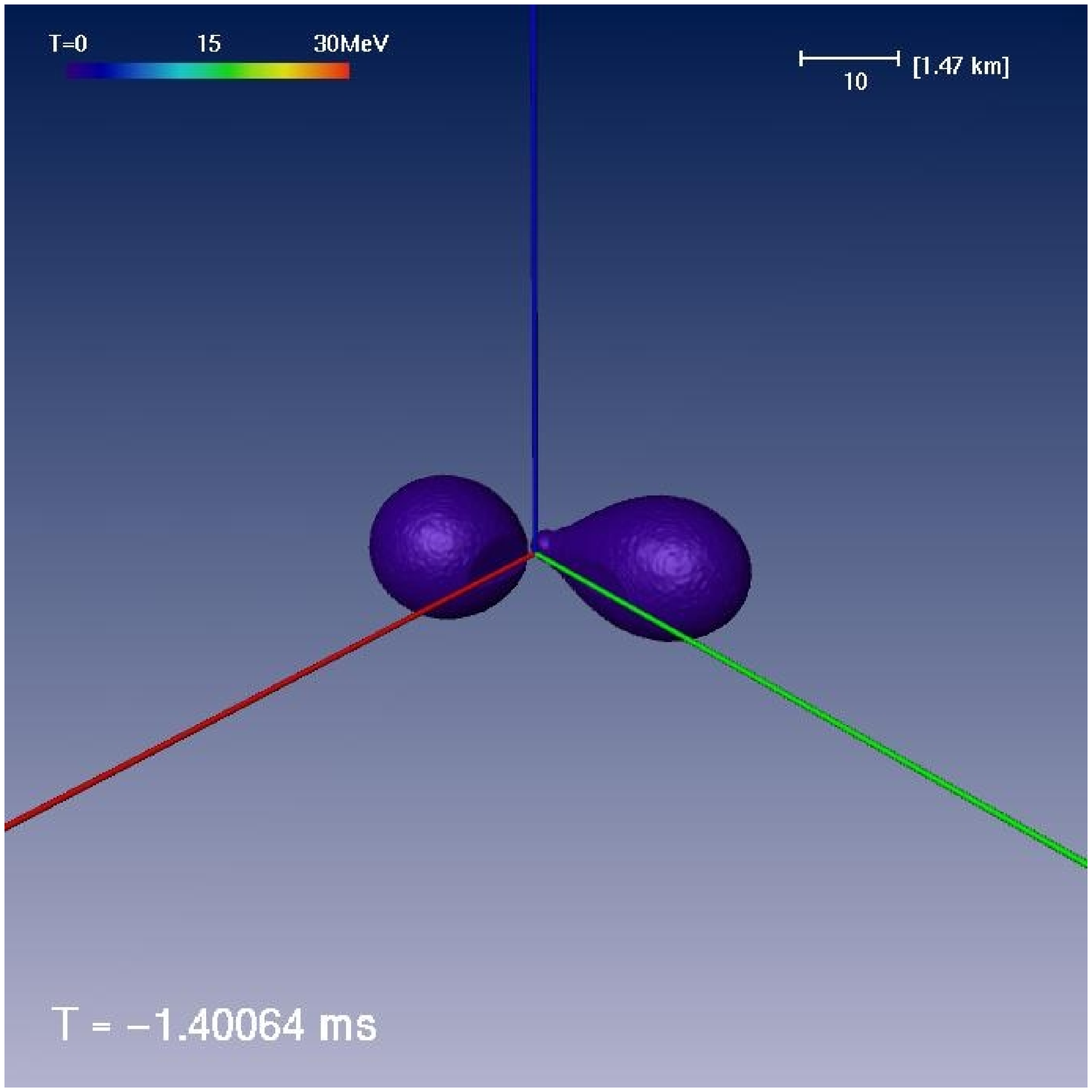}
\end{minipage}
\begin{minipage}[t]{0.47\linewidth}
        (b)
   \includegraphics[width=6.7cm]{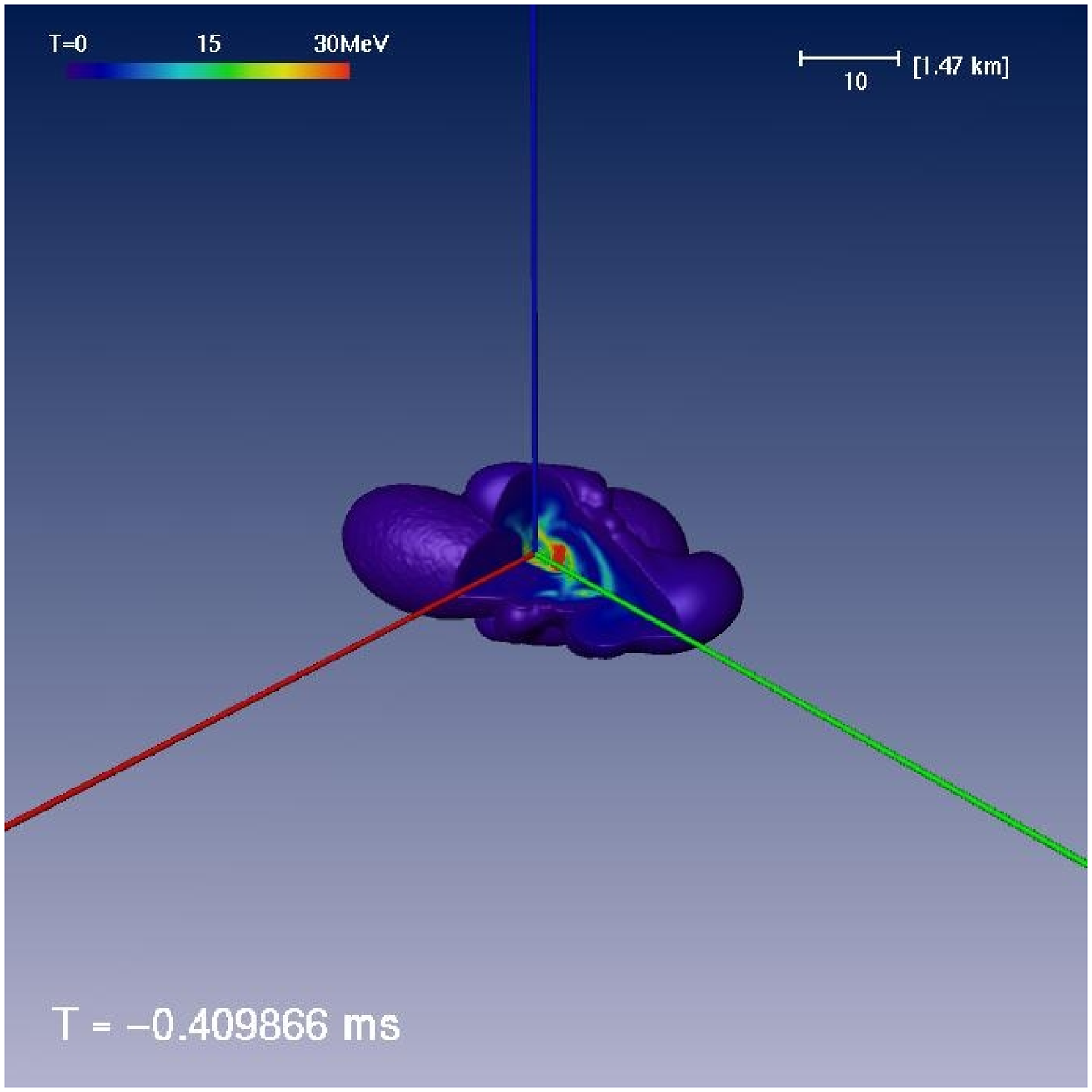}
\end{minipage}
\begin{minipage}[t]{0.47\linewidth}
        (c)
   \includegraphics[width=6.7cm]{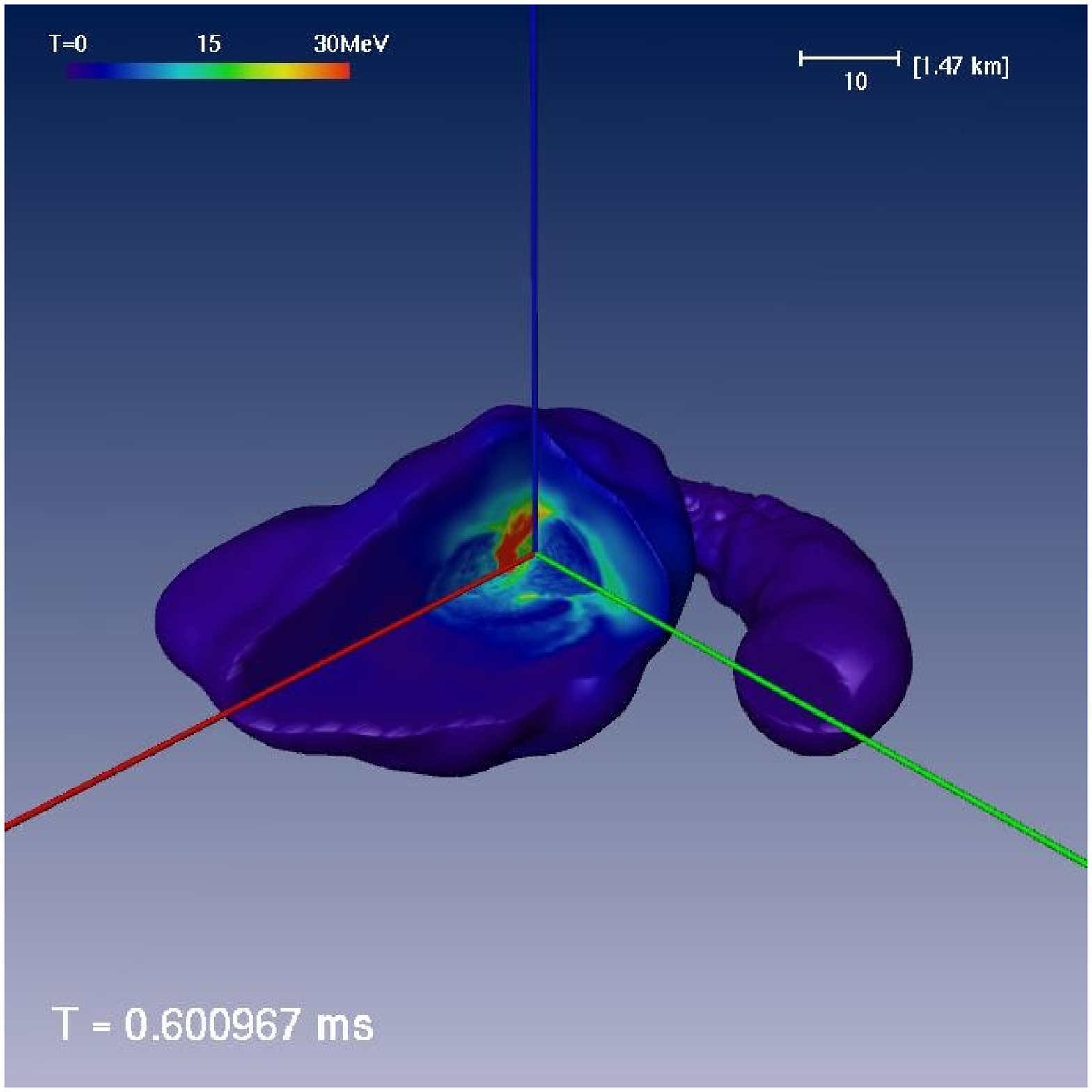}
\end{minipage}
\begin{minipage}[t]{0.47\linewidth}
        (d)
   \includegraphics[width=6.7cm]{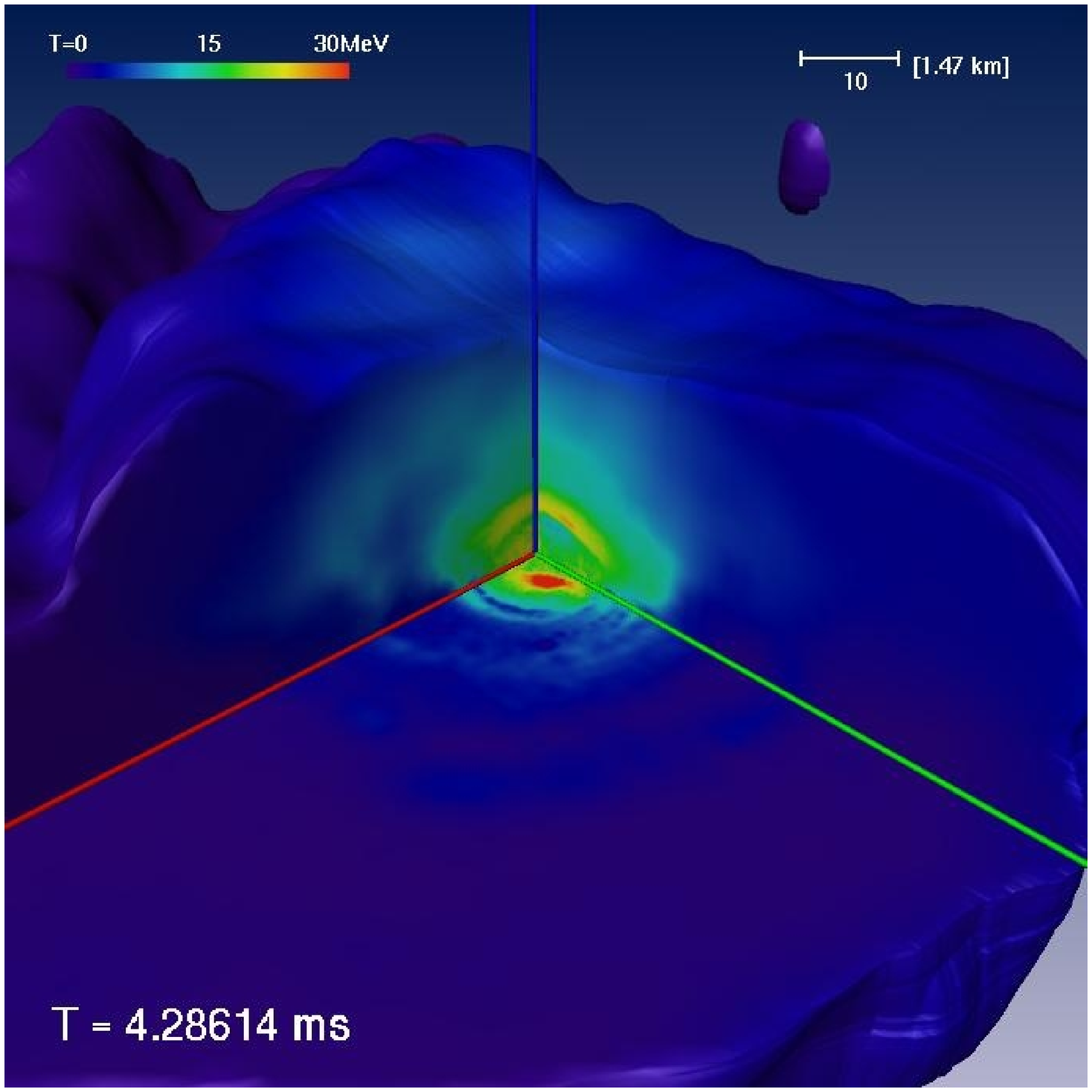}
\end{minipage}
  \caption{Four stages of the NS+NS merger Model S1216. The
surface is chosen to correspond to a density of $10^{10}\,$g$\,$cm$^{-3}$,
the temperature distribution is visible color coded in the octant cut
out from the three-dimensional mass distribution. Temperatures up to
30$\,$MeV are reached at the center only milliseconds after the two
neutron stars have merged. Time is given in the lower left corner of
the panels, and length is measured in units of 1.47$\,$km (top right
corner of each panel).}
 \label{fig:temperatures}
\vspace{-0.5cm}
\end{center}
\end{figure*}

\subsection{Constraints from the low-energy, short-duration GRB~050509b}

If further observations substantiate the link between NS+NS mergers 
and short GRBs, this may set constraints on the post-merging 
evolution and the nuclear EoS. On the one hand, the hypermassive
NS should escape the collapse to a BH for a sufficiently
long time so that an accretion torus can form by the described 
dynamical processes. On the other hand, the collapse of the hot, 
neutrino radiating NS should not be delayed too much, because
neutrino energy deposition in the surface-near layers of the NS
will lead to a massive baryonic wind (Duncan et al.\ 1986,
\citealt{woosley1993}, \citealt{qian1996}, \citealt{thompson2001}),
which can seriously endanger the subsequent formation of 
a GRB jet or fireball.

The mass loss rate of the neutrino radiating neutron star due
to this nonrelativistically expanding outflow of bayonic matter
was estimated by \citep{woosley1993} and \citep{qian1996} and 
is approximately given by (see \citealt{heger2005})
\begin{equation}
\dot M_{\mathrm{wind}}\,\approx\, 
8.5\times 10^{-3}\, L_{\nu_e+\bar\nu_e,53}^{5/3}
\, R_{30}^{5/3}\, M_3^{-2} \ M_\odot\,\mathrm{s}^{-1} \ ,
\label{eq:nuwind}
\end{equation}
where $L_{\nu_e+\bar\nu_e,53}$ is the luminosity of electron neutrinos
plus antineutrinos in units of $10^{53}\,$erg$\,$s$^{-1}$, $R_{30}$ the
NS radius normalized to 30$\,$km, and $M_3$ the NS mass in units of    
$3\,M_\odot$. This mass loss rate can be up to
$\sim\,10^{-2}\,M_\odot\,$s$^{-1}$ for the neutrino luminosities 
found in NS+NS merger simulations and the masses
and radii of the compact post-merger object 
(see \citealt{ruffert2001}, \citealt{rosswog2002b}, 
\citealt{rosswog2003}). Upon colliding with each other, the two
merging neutron stars heat up by shocks and compression to
central temperatures of several 10$\,$MeV (Figs. \ref{fig:morph}, \ref{fig:temperatures}).
Within only a few milliseconds after the final plunge, the neutrino 
emission reaches luminosities of $10^{53}\,$ergs$\,$s$^{-1}$ or higher
(for more information, see \citealt{ruffert1997}, \citealt{ruffert2001}, 
\citealt{rosswog2003}, \citealt{rosswog2003c}), and the post-merger
object should start losing mass in the neutrino-driven wind\footnote{The
development of the neutrino-driven wind is not seen in
current NS+NS merger simulations, because it requires a suitable 
treatment of neutrino transport, whereas the present models describe
neutrino effects with a trapping scheme at best, which releases 
neutrinos from the stellar medium according to their local
production rates, properly reduced due to a finite diffusion 
timescale (see \citealt{ruffert1996} and \citealt{rosswog2003}).}. 
Equation~(\ref{eq:nuwind}), however,
does not take into account the effects of the very rapid rotation of the
neutron star on the wind. This question is still
unexplored, but the rapid spinning of the remnant born in a NS+NS merger 
event may lead to changes of the functional dependence 
of $\dot M$ on the NS properties and may also imply a strong 
pole-to-equator difference of the mass-loss rate.

\begin{table}
\begin{center}
\caption{Estimated accreted masses, $M_{\mathrm{acc}}$, and lower
bounds for the average mass accretion rates,
$\dot M_{\mathrm{acc}}$, assuming post-merger BH-torus systems
being the energy sources of the recently observed, well-localized short GRBs.
The estimates are based on Eqs.~(\ref{eq:torusmass}) and (\ref{eq:accrate}).
The quantity $E_{\gamma,{\mathrm{iso}}}$ is the isotropic-equivalent
$\gamma$-ray burst energy (corrected for the cosmological redshift $z$ of
the burst), and $t_\gamma$ is the GRB duration at the source, computed
as $t_\gamma = T_{90}/(1+z)$ from the measured 90\%-inclusive interval of
high-energy emission. The observational data were taken from \citet{fox2005}.}
\label{tab:discmasses}
\begin{tabular}{l|c|l|c|c|l}
GRB & $z$ & $t_\gamma $ & $E_{\gamma,{\mathrm{iso}}}$
    & $M_{\mathrm{acc}}$ & $\dot M_{\mathrm{acc}}$ \\
\hline
Unit & $\phantom{\displaystyle{\frac{M_\odot^2}{M_\odot}}}$  & sec. & ergs &
$M_\odot$ & $\displaystyle{\frac{M_\odot}{\mathrm{s}}}$ \\
\hline
050509b & 0.225 & 0.033 & $4.5\times 10^{48}$ & $2.5\times 10^{-3}$ & 0.08 \\
050709  & 0.160 & 0.060 & $6.9\times 10^{49}$ & $3.8\times 10^{-2}$ & 0.6  \\
050724  & 0.258 & 2.4   & $4.0\times 10^{50}$ & $2.2\times 10^{-1}$ & 0.09 \\
050813  & 0.722 & 0.35  & $6.5\times 10^{50}$ & $3.6\times 10^{-1}$ & 1.0  \\
\end{tabular}
\end{center}
\end{table}

The nonrelativistic wind will create an extended baryonic halo
that surrounds the merger remnant, before the
compact central part undergoes collapse to a BH and the outer,
high-angular momentum matter assembles in a torus around the
BH. A jet launched from such a system and expanding into the
pre-existing wind halo may then sweep up so much mass -- even
if the halo has a low density and small mass -- that the high Lorentz
factor outflow needed for GRBs ($\Gamma \ga 100$) is prevented.
This was, for example, seen in the type-A models of \citet{aloy2005}.
A rather fast collapse
of the relic hot, supramassive or hypermassive remnant to a BH 
is therefore required, if the cloud of baryonic matter produced
by the neutrino-driven wind from the transiently stable NS
shall not become such a hazard for the GRB\footnote{Note that  
neutrino energy deposition will lateron also drive a baryonic
wind off the accretion torus, after the latter was heated by
viscous dissipation of rotational energy and has started to 
radiate neutrinos with high luminosities (see the discussion
in \citealt{ruffert1997}, \citealt{ruffert1999},
\citealt{rosswog2002b}, \citealt{rosswog2003b},
and find information about the neutrino emission
from such accretion tori in \citealt{setiawan2004} and
\citealt{lee2005b}, and references therein). This wind
does not provide such a hazard for GRB-viable outflows, because
it originates from outside the innermost stable circular
orbit, and angular momentum conservation prevents it from
filling the axial regions above the poles of the BH. In fact,
 \citet{rosswog2003b} and \citet{rosswog2003c} 
consider the pressure of such a ``wind envelope'' as helpful 
for the collimation of a $\nu\bar\nu$-annihilation-driven axial
jet (see \citealt{levinson2000}), an effect which turned out 
to be secondary for getting collimated, highly-relativistic
outflow from BH-torus systems in recent hydrodynamical jet 
simulations \citep{aloy2005}.}.

A corresponding mass limit for the wind halo around the newly 
formed BH-torus system can be derived by the following arguments.
In order for the baryon loading to remain sufficiently low so
that acceleration to Lorentz factors above a value $\Gamma_0$ 
(100 or more) is possible, the mass of the halo swept up by two
relativistic GRB jets, $M_{\mathrm{swept}}$, is constrained by the condition
$E_{\mathrm{kin}}(\Gamma > \Gamma_0)/(M_{\mathrm{swept}}c^2) > \Gamma_0$.
Here $E_{\mathrm{kin}}(\Gamma > \Gamma_0)$ is the kinetic energy of
matter which can accelerate to Lorentz factors larger than 
$\Gamma_0$ in the two polar jets
that expand away from the BH in the axial direction. For a jet collimation
factor $f_\Omega$, this kinetic energy
is linked to the isotropic equivalent $\gamma$-energy of the GRB
by $E_{\mathrm{kin}} = E_{\gamma,\mathrm{iso}}f_\Omega/f_4$, when 
$f_4$ is again
the efficiency for the conversion of kinetic energy to gamma rays.
On the other hand, the wind mass swept-up by the jets is given as
$M_{\mathrm{swept}} = f_\Omega\alpha M_{\mathrm{wind}}$, where 
$M_{\mathrm{wind}} \approx \dot M_{\mathrm{wind}}t_{\mathrm{NS}}$ is the 
mass of the wind for a lifetime $t_{\mathrm{NS}}$ of the neutron star,
and $\alpha < 1$ is a fudge factor which accounts for the fact
that the wind from a rapidly rotating neutron star might be less dense
along the rotation axis than calculated from Eq.~(\ref{eq:nuwind}).
Combining all, we get the condition
\begin{equation}
t_{\mathrm{NS}} \,\la\, {E_{\gamma,\mathrm{iso}} \over \Gamma_0\,f_4\
\alpha\,\dot M_{\mathrm{wind}}\,c^2} \ .
\label{eq:nslifetime}
\end{equation}
With $\Gamma_0 = 100$, $f_4\sim 0.1$, $\dot M_{\mathrm{wind}} =
10^{-3} M_\odot\,$s$^{-1}$ (this assumption is on the low side of
reasonable numbers from Eq.~\ref{eq:nuwind}), and 
$E_{\gamma,\mathrm{iso}}$ from
Table~\ref{tab:discmasses}, we find for GRB~050509b (the lowest-energy
case and thus the one providing the strongest limits) that 
$t_{\mathrm{NS}} < 1\,$ms, assuming $\alpha = 1$.
This means that due to the low energy of GRB~050509b even a tiny 
amount of baryonic pollution by swept-up wind matter would have
prevented ultrarelativistic motion of the jets. The neutrino 
radiating merger remnant therefore must have collapsed
to a BH essentially immediately after the binary NS merging and
the onset of the neutrino-driven wind.
Since it cannot be excluded that the density of the neutrino-driven 
wind along the rotation axis is much lower than estimated by 
Eq.~(\ref{eq:nuwind}), a more conservative estimate with
$\alpha \sim 0.01$ yields $t_{\mathrm{NS}} < 100\,$ms.

If GRB~050509b originated from a NS+NS merger, this result 
therefore suggests that the merger remnant was a hypermassive
object, which was transiently 
stabilized by its rapid and differential rotation 
\citep{baumgarte2000}, but had a very short lifetime.
Referring to the analysis by \citet{morrison2004} -- see their
Table 2 -- this can either mean (i) that, if the high-density EoS
was stiff, the system mass of the merging binary was  
unusually high compared to the known galactic double neutron
stars, which all possess combined masses in the range of 
$M_{\mathrm{sum}} = M_1+M_2 = 2.6$--2.8$\,M_\odot$ \citep{stairs2004};
or, alternatively, (ii) that the nuclear equation of
state is so soft that it could not support the 
merger remnant despite of its rapid differential rotation and
thermal pressure; or (iii) that a very efficient
mechanism was at work which destroyed the differential rotation of the
compact remnant and extracted a sizable amount of its angular momentum
on the required short timescale. This could, for example, happen by
hydrodynamical effects and mass shedding and/or by strong
gravitational wave emission (\citealt{shibata2005}). 

Case (i) might be rejected as unlikely. Case (ii) tends to favor the
``softest'' of the EoSs surveyed by \citet{morrison2004} which 
support nonrotating NSs only up to a gravitational mass around
1.65$\,M_\odot$. Such EoSs, however, are excluded by the recent
accurate measurement of a mass of 2.1$\,M_\odot$ for the millisecond
pulsar PSR$\,$J0751+1807 \citep{nice2005} and by other neutron 
stars with masses beyond $2\,M_\odot$ (see, e.g., the review by \citealt{lattimer2004}). An EoS which is sufficiently stiff 
to fulfill this observational constraint can also ensure
the transient stability of the merger remnant of binary NS systems 
with a canonical mass of 2.6--2.8$\,M_\odot$ (see the
discussion of the critical mass for hypermassive NSs by
\citealt{shibata2005b}). Therefore case (iii) seems to offer a
plausible scenario for GRB~050509b. The NS+NS binary had a typical
mass and the merger remnant was well
above the maximum mass of stable, rigidly rotating neutron stars.
The redistribution of angular momentum by non-axisymmetric  
hydrodynamic interaction and gravitational radiation
(\citealt{shibata2005}) or by viscosity and MHD effects 
(\citealt{shapiro2000}) has then driven the hypermassive object to
gravitational instability on a timescale of $\ll\,$100$\,$ms.
The short accretion timescale ($t_{\mathrm{acc}} < t_\gamma$)
suggests that the remaining torus mass was small. This is consistent
with our conclusion that little mass was accreted by the BH, 
which we have drawn from the low energy of GRB~050509b 
(Table~\ref{tab:discmasses}). And it is consistent with
the requirement coming from optical limits that little radiating
material was ejected (\citealt{hjorth2005a}), e.g. by neutrino-driven
winds or by mass shedding due to viscous transport of angular 
momentum within the accretion torus. In view of the torus
formation models presented here, this points to a 
symmetric or nearly symmetric NS+NS binary with a typical
mass as the source of GRB~050509b.

\section{Discussion}

Although our analysis of the four well-localized short GRBs
was based on a number of unsettled assumptions, we found that
the measured GRB energies and durations lead to estimates
for $M_{\mathrm{acc}}$ and $\dot M_{\mathrm{acc}}$ that are
roughly consistent with expectations from theoretical models
of NS merging and post-merging evolution.
However, in spite of this amazing result 
we think that it would be premature to
claim that relic BH-torus systems from binary NS mergers and 
their associated neutrino emission can power all short GRBs.

Short GRBs have durations between a few milliseconds and about
two seconds, and they might differ in their energy output by 
a factor of 100 or more. Observations also reveal large differences
with respect to their lightcurves and spectral hardness ratios.
This wide range of short GRB properties leaves plenty of room
for the possibility that different compact binaries contribute
to the observed events. For example, not only NS+NS but also 
NS+BH mergers can lead to BH-torus systems with sizable torus
masses and favorable properties for GRB production.
Even the physical processes which play a dominant role in the
energy release of the central engine and jet formation 
might differ between the events. In fact, it might turn out
that this is needed to account for the observed diversity
of short GRBs and their afterglows. Flare-like events 
followed by a very rapid decay were, for example, observed in
case of GRB~050724 hundreds of seconds after the initial GRB.
This was interpreted as a consequence of long-time 
activity of the central engine, which might point to a partial
disruption and gradual accretion of a NS by a BH 
\citep{barthelmy2005}. Episodic and long-lasting mass transfer
from the NS to the BH was found to be possible 
for certain NS/BH mass ratios in combination with special
properties of the NS equation of state (see e.g., 
\citealt{kluzniak1998}, \citealt{portegies1998}, \citealt{lee1999},
\citealt{janka1999}, \citealt{rosswog2004}, \citealt{rosswog2005}, \citealt{davies2005})\footnote{Simulations 
and analytic studies of NS+BH
mergers have so far been carried out only in Newtonian
gravity and disregarding the probable rotation of the BH.
A Newtonian treatment, however, does not reproduce the
correct mass-radius relation of a relativistic neutron star,
for example.
Current predictions of the merger evolution and of the
dynamics of disc formation and accretion in such extremely 
relativistic
systems can therefore not be considered as reliable and
conclusive, neither quantitatively nor qualitatively.}.



The rather uncertain estimates of the merger
rates also allow for the possibility that not all NS+NS/BH mergers 
produce GRBs (see, e.g., \citealt{guetta2005}, taking into 
account that the hydrodynamic models by \citet{aloy2005}
yield jet semi-opening angles of $\sim\,$10$^{\mathrm{o}}$
instead of a value of $\sim\,$1$^{\mathrm{o}}$ as found by 
Guetta \& Piran with the assumption that all mergers make GRBs).

We therefore refrain (unlike others, see, e.g., \citealt{lee2005},
\citealt{rosswog2003c}) from making predictions
of the distribution of GRB properties on grounds of current 
simulations of NS+NS or NS+BH mergers. Such attempts are
seriously hampered by the large number of unknowns in the 
theoretical picture and in particular their relation to the 
properties of compact object mergers:
\begin{itemize}
\item    The distributions of binary parameters 
(NS and BH masses and spins, binary mass ratios) are highly uncertain. 
Theoretically, because of our incomplete understanding of the binary
formation and 
evolution and of the supernova explosions which terminate the lives
of massive stars and give birth to neutron stars \citep{bulik2003}.
And observationally, because of the rather
limited sample of known NS+NS or NS+BH binaries (e.g., \citealt{lattimer2004}, \citealt{stairs2004}). The latter aspect
is particularly bothersome in view of the fact that only a minor
fraction (of order 10\%?) of NS+NS/BH mergers might be able to
produce GRBs and might be needed to account for the observed short
bursts.
\item    NS+NS/BH merger models require much more
work. Relativistic simulations of compact object mergers are
needed for different (non-zero temperature) nuclear equations 
of state and different parameters of the binary systems. 
A softer nuclear EoS, for example, might lead to a relation between 
binary parameters and torus masses that differs from the results
of this work, which are based on the rather stiff EoS of
\citet{shen1998, shen1998b}. NS+NS merger models must be consistently evolved
from the last stages of the progenitor system, through the moment of BH 
formation, to the subsequent accretion and jet production phase,
including all the potentially relevant physics like neutrino 
transport and magnetic fields. Only then, for example, will the 
question be answered under which circumstances
the neutrino-driven wind from the transiently stable hypermassive
merger remnant represents an unsurmountable obstacle for GRB jets.
\item    The complex sequence of physical processes from the energy
release by the merger remnant to the stage of GRB production 
involves large uncertainties, expressed by the product of 
efficiency factors $f_1$ to $f_4$ and jet collimation factor 
$f_\Omega$, which appears in
Eq.~(\ref{eq:torusmass}). Each of these factors can contain 
still unknown dependences on the parameters of the NS+NS/BH binary 
and of its relic BH-torus system. Improved simulations of 
jet formation by energy release around post-merger BH-torus
systems are necessary to clarify these dependences.
\item    Since the compact proto-BHs found in our NS+NS merger
simulations have significant angular momentum 
($a_{\mathrm{rem}}$ is between 0.64 and 0.91 for the 
considered mergers of irrotational binaries; Table~\ref{tab:inittable}),
the energy release during the accretion is likely to be boosted
by Kerr effects. Not only neutrino emission and neutrino-antineutrino
annihilation may play a role, but also energy release through 
magnetic processes, e.g., by the Blandford-Znajek (1977) mechanism 
(see, for example, \citealt{lee2005}).
\end{itemize}
Much more theoretical work and numerical modeling are therefore 
needed before
truly meaningful and reliable theoretical predictions of  
GRB properties and their link to NS+NS/BH merger parameters will
become possible. Such predictions need to invoke,
in particular, detailed models of the structure of the collimated   
ultrarelativistic outflows from the merger remnants and assumptions
about the production of the GRB emission at large distances.
Detailed consideration of the relevant physics is clearly
beyond the scope of this paper.
A first, still very crude attempt to employ such calculations of
jets from BH-torus systems for predicting the distributions of
observable GRB properties has been attempted recently by
\citet{janka2005}. 

\section{Summary and conclusions}

We have shown that nonradial oscillations and triaxial
deformation and the associated tidal torques can mediate
efficient angular momentum transfer
in the compact remnant within the first milliseconds after
the merging of two NSs. Applying different criteria,
we estimated the mass of the torus which will survive the
future collapse of the central, dense object to a BH. For the
considered irrotational systems and the employed relatively
stiff nuclear EoS of \citet{shen1998,shen1998b}, we
determined lower disc mass limits of 1 -- 2\% of the (baryonic)
system mass for a NS-NS (gravitational) mass ratio around $q = 1$,
increasing to about
9\% for systems with $q$ around 0.55.
Though the disc mass is more sensitive to the parameter $q$,
we also observed a mild inverse relation between
torus mass and binary system mass. For the same value of $q$,
slightly higher torus masses were found when $M_{\mathrm{sum}} =
M_1 + M_2$ was lower (Fig.~\ref{fig:discmasses_summary}).
Non-zero temperature effects turned out to be
important and to {\em reduce} the disc mass by several 10\%
compared to the values for the $T=0$ case.

The disc masses obtained in our simulations confirm the
viability of post-merging BH-torus systems as central engines
of short GRBs, and they support the theoretical possibility that
thermal energy deposition above the
poles of the BH by $\nu\bar\nu$ annihilation could be the
primary energy source of short GRBs. The ultrarelativistic
outflow of matter powered by this energy deposition was
found in hydrodynamic simulations to be highly collimated in jets
with semi-opening angles between about 10$^{\mathrm{o}}$ and
15$^{\mathrm{o}}$ \citep{aloy2005}. This
prediction seems to be supported
by the observations of two of the four recently detected
well-localized, short hard GRBs \citep{fox2005}.

Using typical parameters that describe the multi-step process from
the energy release of the central engine to the observed GRB in
this scenario, we derived estimates of the accreted masses and
mass accretion rates (see Table~\ref{tab:discmasses}) from
the measured GRB energies and durations, using
Eqs.~(\ref{eq:torusmass}) and (\ref{eq:accrate}).
For all cases we found values for $M_{\mathrm{acc}}$
and $\dot M_{\mathrm{acc}}$ that are in the ballpark of
the expectations from our NS+NS merger models and from models of
BH accretion in post-merging BH-torus systems.
Employing the same assumptions about the GRB engine,
the extraordinarily low-energetic and short-duration GRB~050509b can
be used to set strict limits for the lifetime of the post-merger
hypermassive neutron star. It must have collapsed to a BH in
much less than 100$\,$ms with only little matter ($\sim 0.01\,M_\odot$)
remaining in a torus around the BH to partially produce GRB-viable outflow. This favors a symmetric or
nearly symmetric NS+NS binary as progenitor system, with a mass
typical of the known galactic double neutron stars.

Knowledge of the
binary system parameters, which could be obtained from measurements
of the gravitational wave chirp signal during the inspiral, might allow
one to deduce interesting information about the properties of the
neutron star equation of state from future short GRBs with
determined distances.
Constraining the EoS of NS matter by lifetime arguments for the
supramassive or hypermassive remnant of NS+NS mergers, however,
will require detailed relativistic models of the NS merging and
of the post-merging evolution for different non-zero temperature
EoSs, including the effects of magnetic fields \citep{shibata2005b}.
GRB~050509b
provides particularly strict limits in such an analysis, but
may be a rare event where an exceptionally low-energy GRB with very
short duration could be well localized.

\section*{Acknowledgments}
We thank M.A.\ Aloy for inspiring discussions and A.\ Marek for
preparing the EoS table used in this work.
Support from the Sonderforschungsbereich-Transregio~7
of the Deutsche Forschungsgemeinschaft is acknowledged.
The computations were performed at the Rechenzentrum Garching.

\bibliographystyle{../mn2e}
\bibliography{../discpaper/biblio}

\end{document}